\documentclass[%
reprint,
nofootinbib,
amsmath,amssymb,
aps,showkeys,showpacs,
prd,
]{revtex4-1}
\usepackage{newtxmath,newtxtext}
\usepackage{graphicx}
\usepackage[caption=false]{subfig}
\usepackage{bm}
\usepackage{dsfont}
\usepackage{amsmath,amssymb,stackengine}
\usepackage{siunitx}
\usepackage[caption=false]{subfig}
\usepackage{braket}
\usepackage{listings}
\usepackage{xcolor}
\usepackage{hyperref}
\usepackage{tcolorbox}
\usepackage{braket}
\hypersetup{colorlinks=true,linkcolor=blue,urlcolor=blue,citecolor=blue}
\usepackage{graphicx}
\usepackage{dcolumn}
\usepackage{bm}

\usepackage{orcidlink}
\usepackage{multirow, makecell}
\usepackage{rotating}
\begin{document}


\title{Note on Granda -- Oliveros Holographic Dark Energy}

\author{Manosh T. Manoharan\orcidlink{0000-0001-8455-6951}}
\email{tm.manosh@gmail.com; tm.manosh@cusat.ac.in}
\affiliation{Department of Physics, Cochin University of Science and Technology, Kochi--682022, India.}%

\date{\today}

\begin{abstract} 
To address the issues surrounding the choice of the infrared (IR) cut-off in standard holographic dark energy (HDE) models, Granda and Oliveros introduced a comprehensive solution widely known as the Granda-Oliveros (GO) cut-off. We refer to this as the Granda-Oliveros HDE (GOHDE) model. This article revisits this model, explores the features of the parameter space and shows that its ability to explain the late-time acceleration is only due to the integration constant that appears by construction. We show that the GOHDE behaves like the dominant energy component and naturally behaves like dark energy in the late phase. In the matter-dominated period, it acts like pressureless matter and exhibits radiation-like behaviour in the very early stages. Interestingly, adjusting the free parameters could subdue or enhance these characteristics. Consequently, we see that the holographic principle applies to all cosmic elements, extending beyond the description of dark energy alone. Depending on the parameters, GOHDE can act similarly to concordance, phantom or quintessence-like dark energies with singular equation of states. Further, from the point of local observations, we show that the GOHDE model is observationally indistinguishable from $w$CDM and encompasses $\Lambda$CDM as a specific case. Our analysis reveals that while a departure from $\Lambda$CDM could account for late-time acceleration, it falls short of a consistent description of the entire cosmic history. As we observe dark energy could behave like pressure-less matter in the past, it brings tension in baryon density, power spectrum, etc. Furthermore, we utilize various datasets, including OHD, Pantheon, CMB Shift parameter, BAO, and QSO, to constrain the free parameters of the GOHDE model. Our analysis indicates that the best fit, assuming the GO cut-off, aligns with the $\Lambda$CDM model. Additionally, we use statistical quantifiers such as AIC, BIC, and $\chi^2$ to test the model's goodness with various data combinations and estimate the Bayes factor to contrast the models. Our study suggests that the standard GOHDE model is equally likely as the $\Lambda$CDM model, ultimately favouring the latter, with $\beta=0.686^{+0.048}_{-0.044}$ and $w{z_0}=-0.988^{+0.042}_{-0.044}$. Given the current observations, we conclude that, for a flat universe, the $\Lambda$CDM case is statistically the best and probably the only consistent solution from the GOHDE construction. In effect, our study questions the foundations of the HDE approach by insisting on a shift in the standard holographic paradigm if we were to address the cosmological constant problems.

\end{abstract}

\keywords{Granda-Oliveros IR cut-off, Cosmological Constant Problem, Holographic Dark Energy}
\maketitle


\section{Introduction \label{sec:1}}
The late-time accelerated expansion of the universe is a well-established observation \cite{Riess_1998, Perlmutter1998, Perlmutter_1997, Perlmutter_1999}, and since this discovery, many debates have been over what drives this accelerated expansion. The simplest and arguably the most elegant explanation that fits remarkably with observation is the existence of a constant energy density, which the concordance $\Lambda$ cold dark matter ($\Lambda$CDM) model assumes as the cosmological constant ($\Lambda$). The fact that it takes only an additional constant to explain late-time acceleration simultaneously makes it look simple and yet challenging. Whatever explains this acceleration goes by the name ``dark energy'' in the literature, and its source is more or less obscure. The obvious candidate from quantum field theory (QFT) that fits the attributes of a cosmological constant is the vacuum energy density, which, in principle, can be extremely large depending on the energy scale we assume QFT's validity. Although the freedom to add a bare constant to the action in Einstein's gravity allows us to cancel out this immense value, it mandates unnatural fine-tuning. This observation leads us to the old cosmological problem or the fine-tuning problem \cite{RevModPhys.61.1, RevModPhys.75.559}. The prediction\footnote{ The vacuum energy density calculated from QFT is not strictly a prediction. If it were, any mismatch would rule out the theory, which is not the case. } from QFT is questionable as it does not account for any effects of gravity. Furthermore, the choice of UV cut-off cannot be strictly the Planck energy, as QFT has not yet been tested in those ranges. The problem deepens when we have no clue from the observations beyond the fact that it is almost a constant. This constancy brings out the second phase of the concern: why is it a constant? Why does it have the value it does? Especially now. These questions aggregate to form the coincidence problem, which has been addressed from various perspectives \cite{PhysRevLett.82.896}. All these arguments also question our preconceptions about a simple constant that explains late-time accelerated expansion \cite{bianchi2010prejudices}. This article reassesses the holographic dark energy paradigm and describes why it accounts for the late time acceleration and cannot solve the cosmological constant problems.

It is well known that the vacuum energy shows both UV and IR divergence even for a simple massless scalar field \cite{Schwartz:2014sze}. One can eliminate the IR divergence by evaluating the energy density instead of the total energy and resolve the UV divergence by setting a proper UV cut-off. However, this regularisation does not incorporate any assumption about the spacetime geometry where the quantum field exists. One can instead consider QFT in curved spacetime with or without any coupling and still show that there is the UV divergence in the asymptotic de Sitter limit \cite{birrell_davies_1982, PhysRevD.84.044040}.   One potential solution to this UV divergence comes from the field of black hole thermodynamics. This solution is often associated with the entropy bounds or, in popular terms, it is known as a holographic solution. The term "holography" is commonly used in the context of dimensional reduction, and in this context, it suggests that the maximum possible entropy of a system depends on its boundary instead of its bulk.

By considering the maximum possible entropy value in a given space, it becomes feasible to establish a limit on the maximum energy within that region. This criterion connects the ultraviolet (UV) and infrared (IR) cut-offs through the boundary entropy. This vision leads to the calculations in \cite{PhysRevLett.82.4971}, which proposes an effective value for the total energy density as,
\begin{equation}
\Lambda^4\sim\rho\lesssim\frac{S}{L^4}.
\label{eq:CKNBound}
\end{equation}
Frequently referred to as the CKN bound, which takes its name from its authors Cohen, Kaplan, and Nelson, it yields an energy density consistent with observational data when applied in a universe \textit{dominated} by dark energy. As a result, according to reference \cite{PhysRevLett.82.4971}, it claims to address the fine-tuning problem effectively. A notable consequence of this endeavour is introducing the concept of holographic dark energy (HDE) \cite{WANG20171}. Their objective was to extend the results presented in \cite{PhysRevLett.82.4971} to settle the coincidence problem and comprehensively resolve the cosmological constant problems. Despite numerous efforts to improve this holographic dark energy approach, there is a need for greater consistency in several aspects. Nevertheless, one can appreciate this approach for its adherence to the fundamental principles, stemming from the investigations of the holographic principle \cite{hooft1993dimensional, 10.1063/1.531249}, which includes the renowned AdS/CFT correspondence \cite{Maldacena1999}.

The notable historical evolution of the HDE models unfolds as follows. By adopting the CKN bound as the expression for dark energy, up to a proportionality constant, Hsu identified that the relationship yields an incorrect equation of state for the dark energy \cite{HSU200413}. They demonstrated that the density might correspond to an alternative form of diffused dark matter. However, the manuscript laid the groundwork for the emergence of the HDE paradigm, a term that became popular following Li's contribution \cite{LI20041}. In contrast to CKN's original proposal of using the Hubble parameter as the IR cut-off, Li considered both the particle and event horizons as the IR cut-offs. However, the particle horizon encountered the same issues as the Hubble horizon, as highlighted by Hsu \cite{HSU200413}. With the event horizon as the IR cut-off, it is possible to explain the recent acceleration with an appropriate dark energy equation of state. Subsequently, work by Myung identified several inconsistencies in Li's constructions \cite{MYUNG2007247}. Myung suggested that the equation of state could be $`-1'$ by considering the dual quantum system of a singularity-free de Sitter black hole \cite{MYUNG2007247}. The author conceded that the correct equation of state of dark energy is more of a consistency condition than a derived result. These results leave room for modifications, and various attempts have been undertaken to create a more robust version of loophole-free HDE. Such attempts include exploring non-standard interactions between the dark sectors \cite{PAVON2005206} and considering non-extensive entropy alternatives in place of the Bekenstein-Hawking area law \cite{PhysRevD.105.044042}, among others.

To resolve the challenges posed by the standard HDE paradigm, Granda and Oliveros introduced a comprehensive IR cut-off, including derivatives of the Hubble parameter \cite{GRANDA2008275}, hereafter the GO cut-off ($L_{\text{GO}}$). The notion of Ricci HDE aligns with this effort, with the notable distinction of containing curvature terms in the Ricci scalar \cite{PhysRevD.79.103509, PhysRevD.79.043511, doi:10.1142/S0217732316500759,doi:10.1142/S0218271819500603}. Several versions of HDE using GO cut-off have been proposed and scrutinised in the literature \cite{Oliveros2022, KOUSSOUR2022115738, Kaur_2023, doi:10.1142/S0218271822501073, Dheepika2022a}. The GO cut-off is claimed to offer several advantages over other alternatives, primarily due to the inclusion of the derivatives of the Hubble parameter, which ensure causality and can explain the late-time acceleration.

While the GO cut-off provides a consistent causal explanation for late-time acceleration, there are reservations regarding its parameter space and construction. The most striking feature of Ricci and GO cut-offs is the presence of derivatives in the definition of dark energy density, making the first Friedman equation a differential equation by construction. The solution of such an expression will always contain an integration constant whose value needs to be fixed by some initial conditions. It is then apparent that this constant drives the late-time acceleration subjected to the values of other free parameters. Although it resembles unimodular theories of gravity, where a cosmological constant appears as the integration constant \cite{universe9030131}, unimodular approaches have a constant arising at the level of field equation. In modified gravity theories, we can set this integration constant to zero and still explain late-time acceleration \cite{zangeneh2023modified}. One must remember that the dynamics associated with any of these constructions depend on the continuity expression, which in the standard HDE paradigm is the same as in the concordance model. Another intriguing feature of Ricci and GO cut-offs is that both give us a dark energy that behaves like pressure-less fluid in the matter-dominated era. We show that it is not a peculiar feature of the IR cut-off but rather a feature associated with the choice of free parameters. We show that these HDE models are indistinguishable from $w$CDM models, and the nature of dark energy is dictated purely by construction. When we survey the parameter space, more interesting behaviours and singular equation of states become apparent in these constructions. While ignoring radiation density or choosing a specific set of free parameters, a notable observation is its ability to retrieve the $\Lambda$CDM model as a particular case. This observation raises intriguing questions: How can we recover $\Lambda$CDM from a model designed to address its shortcomings? Does the construction of HDE presuppose certain aspects that have gone unnoticed until now? Furthermore, we must reconsider how the HDE construction or the CKN bound effectively resolves the fine-tuning problem. This manuscript will address these concerns systematically by deriving analytical expressions and observational constraints. Using the Markov chain Monte Carlo (MCMC) method and various data sets, we will impose restrictions on the parameter space and evaluate the model's statistical significance using multiple quantifiers.

The article's structure is as follows. In the next section, we introduce the GOHDE model and the HDE paradigm in general. Subsequently, we will derive crucial cosmic variables that enable us to probe the characteristics of this GOHDE construction. We then thoroughly explore each variable and its properties within the parameter space. Following this, we perform the data analysis utilizing available observational datasets, leading to a discussion of the nature of the best-fit values. This analysis is followed by various statistical tests to compare the resilience of the models, and then we summarize.

\section{Holographic paradigm \& Infrared cut-off\label{sec:2}}

Many authors have explored several distinct methods in the quest to understand the late-time acceleration of the universe. Our central focus is not centred on developing a gravitational theory capable of explaining accelerated expansion without invoking the concept of dark energy, as explored in \cite{harada2023gravity, PhysRevD.76.103516}. Therefore, for this and subsequent sections, we assume that a non-interacting dark energy component, represented as $\rho_{\Lambda}$, is inherently present. While it is conceivable that one could potentially reformulate this framework to eliminate the dependence on dark energy, such endeavours are reserved for the future.

For a homogeneous and isotropic universe, we begin with the standard Friedmann -- Lema\^{i}tre -- Robertson -- Walker (FLRW) metric, 
\begin{equation}
ds^2=c^2dt^2-a^2\left(\frac{dr^2}{1-kr^2}+r^2d\theta^2+r^2\sin^2\theta d\phi^2\right). 
\end{equation}
Here, $a\equiv a(t)$ represents the scale factor, $k$ signifies the spatial curvature, and $\left(r,\theta,\phi\right)$ denotes the conventional spherical polar coordinates with $t$ as the cosmic time. For simplicity, one can assume a flat universe by setting $k=0$ and use this metric to derive the left-hand side of the Friedmann equations.

Establishing the energy density profile is imperative to construct the right-hand side of the Einstein field equation. In exploring dark energy, we will deliberately maintain the behaviour of energy densities such as matter or radiation unchanged. In other words, the characteristics of every cosmic component, except dark energy, are determined by established principles of standard physics. Notably, the behaviour of matter or radiation components is contingent upon the knowledge or assumption of their respective barotropic pressure. Strikingly, this vital information is absent in the construction of standard HDE \cite{MYUNG2007247}. The expression for standard HDE density $(\rho_{\Lambda})$ is given as \cite{WANG20171},
\begin{equation}
\rho_{\Lambda}={3C^2M^2_{\text{\tiny P}}L^{-2}}.
\label{SHDE_density}
\end{equation}
Here, $L$ denotes the IR cut-off, $C$ is a constant, and $M_\text{\tiny P}$ represents the reduced Planck mass. The Planck mass comes into the picture because of the choice of units with $\hbar=c=1$, where $8\pi G=1/M^2_{\text{\tiny P}}$. This relationship originates from the CKN bound \cite{PhysRevLett.82.4971}, which for a dark energy-dominated universe establishes a correlation between the UV cut-off ($\rho_{\Lambda}$) and the IR cut-off ($L$) via the constraint imposed by the horizon entropy ($S$), given by $\rho{_\Lambda} L^4\lesssim S$.

With the expression for dark energy density in hand, the conventional approach in HDE involves utilizing the standard Friedmann equations to study the universe. The Friedmann equations and the continuity equation for a 3+1 dimensional spacetime are,
\begin{align}
H^2&=\frac{8\pi G}{3}\rho,\label{eq:FriedmannEq1}\\
\dot{H}+H^2&=\frac{-4\pi G}{3}\left(\rho+3p\right),\label{eq:FriedmannEq2}\\
\dot{\rho}&=-3H\left(\rho+p\right)\label{eq:ContinEq}.
\end{align}
In the above expressions, $\rho$ represents the total energy density, which consists of different components, including matter ($\rho_m$), radiation ($\rho_r$), dark energy ($\rho_\Lambda$), and other cosmic members that are not explicitly mentioned. Meanwhile, $p$ represents pressure, and $H$ denotes the Hubble parameter.

\subsection{Constructing the GOHDE}

When formulating an HDE model using Eq. (\ref{SHDE_density}), there are two primary elements to consider. The first is the boundary entropy, and the second is the infrared (IR) cut-off. Various viable options exist for boundary entropy, ranging from the Bekenstein-Hawking entropy to several other interesting alternatives \cite{NOJIRI2022137189}. Despite variations in the details, all choices for entropy share a common feature, adhering to the area law, as established by many seminal works \cite{Bardeen1973, PhysRevD.15.2738, PhysRevD.13.191, Hawking1975, HAWKING1974}. When it comes to the IR cut-off, there isn't a universally accepted golden rule or prescribed formula. Employing the Hubble scale and particle horizon as the IR cut-off proves inadequate in explaining late-time cosmic acceleration\footnote{CKN bound's original motivation presumed a cosmological constant with an equation of state `$-1$', a detail overlooked in \cite{HSU200413}. The sole deduction from \cite{PhysRevLett.82.4971} is that a universe with a large IR cut-off may have a small cosmological constant.}, in contrast to the future event horizon, which demonstrates potential in addressing this phenomenon \cite{LI20041}. Intriguingly, one can employ the Hubble scale as an IR cut-off and account for late-time acceleration, albeit at the expense of relying on generalized entropies instead of the Bekenstein-Hawking relation \cite{TAVAYEF2018195}. Dealing with the logical challenges associated with employing the future event horizon and concerns about causality may be made feasible by incorporating terms involving derivatives of the Hubble parameter. The most apparent choice in this context is the Ricci curvature, as demonstrated in \cite{PhysRevD.79.103509, PhysRevD.79.043511}.

In pursuit of a more comprehensive approach, Granda and Oliveros introduced a generalized IR cut-off, expressed as \cite{GRANDA2008275},
\begin{equation}
L_{\text{GO}}=\left(\alpha H^2+\beta\dot{H}\right)^{-1/2}.
\end{equation}
With this IR cut-off in place, we can define the dark energy density using Eq. (\ref{SHDE_density}) as,
\begin{equation}
\rho_{\Lambda}=3M^2_{\text{\tiny P}}\left(\alpha H^2+\beta\dot{H}\right).
\label{Standard_GOHDE}
\end{equation}
Here, we've absorbed the constant $C$ into the constant $\alpha$ and $\beta$. We will address the details of Eq. (\ref{Standard_GOHDE}) when discussing the $\Lambda$CDM scenario, and in the meantime, we will consider Eq. (\ref{Standard_GOHDE}) as the Granda -- Oliveros HDE density by construction. An apparent assumption inherent in developing HDE is the choice of the entropy form. To obtain the HDE in the above form, we must assume that $S$ scales as $L^2$. Any modification to the entropy will alter this assumption. Consequently, the Bekenstein-Hawking area law is integrated into the construction of the HDE using Eq. (\ref{SHDE_density}), and the only option is to play with different IR cut-offs. We shall address concerns regarding the horizon entropy definition that went into the construction of HDE later. 

In the standard HDE paradigm, one utilizes Eqs. (\ref{eq:FriedmannEq1}) and (\ref{eq:FriedmannEq2}) to investigate cosmic evolution. When we substitute Eq. (\ref{Standard_GOHDE}) into the first Friedmann equation, we obtain a differential equation of the form,
\begin{equation}
H^2 = \frac{8\pi G}{3}\left[\rho_m+\rho_r-ka^{-2} + 3M^2_{\text{\tiny P}}\left(\alpha H^2 + \beta\dot{H}\right)\right].
\end{equation}
Assuming that both ordinary matter and dark matter exhibit the same gravitational behaviour, we can express the total matter density as $\rho_m = \rho_{m0}(a)^{-3}$. Similarly, the radiation goes as $\sim a^{-4}$ and curvature scales like $\sim a^{-2}$. To simplify further, we can normalize the expression by the present value of the Hubble parameter, denoted as $H_0$, and substitute $8\pi G\rho_{i0}/(3H_0^2)$ with $\Omega_{i0}$. This parametrization allows us to explore the characteristics of the dark energy in terms of a varying dark energy equation of state.

Now, one has the flexibility to manipulate different variables. For those interested in investigating the early phase to the present, the suitable variable would be the scale factor ($a$), ranging from zero to unity. Meanwhile, redshift ($z$) proves beneficial for understanding from the present (including the recent past) to the future. Choosing $x = \ln(a)$ would provide a comprehensive outlook on the past and the future. We will interchange between these variables as needed. These transformation allows us to express $\dot{H}$ as $H(dH/dx)$, and we represent $a$ in terms of $(z)$ as $a=1/(1+z)$ when needed. With these notions, we rewrite the previous equation as,
\begin{equation}
h^2=\Omega_{r0}e^{-4x}+\Omega_{m0}e^{-3x}+\Omega_{k0}e^{-2x}+\alpha h^2+\beta h\frac{dh}{dx}.
\end{equation}
The above relation is a first-order non-linear differential equation in $h$, for which finding an analytical solution might be challenging. However, we could solve it as a first-order ordinary differential equation in $h^2$ with $h^2(x=0)=1$, which we can assume by construction. Thus, solving for $h^2$ we get,
\begin{widetext}
	\begin{align}
	h^2&= \frac{\text{$\Omega_{r0}$}e^{-4 x}}{-\alpha +2 \beta +1}+\frac{2  \text{$\Omega_{m0}$}e^{-3 x}}{-2
		\alpha +3 \beta +2}+\frac{ \text{$\Omega_{k0}$}e^{-2 x}}{-\alpha +\beta +1}+\text{C}_1e^{-\frac{2 (\alpha -1) x}{\beta }}\label{FullHubble},\\
	\text{C}_1& =\frac{\alpha
		+\text{$\Omega_{k0}$}}{\alpha -\beta -1}-\frac{2 \text{$\Omega_{m0}$}}{-2 \alpha +3 \beta +2}+\frac{\text{$\Omega_{r0}$}}{\alpha -2 \beta -1}+\frac{\beta
	}{-\alpha +\beta +1}+\frac{1}{-\alpha +\beta +1}.
	\end{align}
\end{widetext}
 
In late-time cosmology with a flat universe, it's reasonable to disregard the contribution of radiation and set $k=0$. This simplification lets us focus exclusively on matter (comprising cold dark matter and baryonic matter) and dark energy, resulting in $\rho = \rho_m + \rho_\Lambda$. In this context, matter is treated as a pressure-less fluid, with $p_m=0$. With this framework, we can explore the dynamics of non-interacting HDE models. Similarly, an additional term representing the preferred interaction should be incorporated into Eq. (\ref{eq:ContinEq}) to explore interacting HDE models. The difficulties and features of such interacting models are extensively discussed in \cite{PhysRevD.106.043527} and the references therein. We are concerned with a scenario where only matter and dark energy are relevant, aiming to contrast the model with observational data. Then the above expression reduces to,
\begin{equation}
h^2=e^{-\frac{2(\alpha -1) x}{\beta }}\left\lbrace 1 -\frac{2\Omega_{\text{m0}}}{2\alpha -3\beta-2}\left[e^{\frac{x (2 \alpha -3\beta -2)}{\beta }}-1\right]\right\rbrace.
\label{eq:HubbleFlow}
\end{equation}
Given we have the expression for $h(x)$, we could substitute for $h=H/H_0$ and $x=-\ln(1+z)$, and we could explore the model in terms of redshift ($z$) or scale factor ($a$). We will call the cosmological model described by Eq. (\ref{eq:HubbleFlow}) the Granda -- Oliveros holographic dark energy with cold dark matter model or, more simply, the GOHDE model. We will make further transformations to the free parameter and rename it in future sections. 

A few immediate observations from the general expression are noteworthy. Suppose we focus on isolating the behaviour of dark energy by excluding matter, radiation, etc.. In that case, we observe that, by design, dark energy mimics the behaviour of matter, radiation, etc., during the corresponding domination periods. Specific parameter choices allow the removal of certain features. For example, when $\alpha=\beta$, the curvature-like behaviour is automatically eliminated. Similarly, radiation-like characteristics are selectively eradicated when $\alpha=2\beta$, mirroring the Ricci scalar and $\alpha=3\beta/2$ picks out matter-like behaviour. In other cases, say $\alpha=3\beta$, none of the features are removed, and dark energy exhibits all such characteristics. Importantly, distinguishing between entities with the same scaling behaviour from observations using just the Hubble parameter is impossible. In the case of baryons and cold dark matter, both scale as $\sim a^3$ and are considered together. Therefore, distinguishing it through measurements that solely account for scaling becomes infeasible if dark energy also displays a similar behaviour. This situation can lead to scenarios involving negative/positive dark energy and super/sub-critical matter density during the past, which will put tensions in estimated baryon density. See Appendix (\ref{AppendixA}) for illustrations.

Another observation is that when $\alpha=1$, the model is virtually indistinguishable from $\Lambda$CDM; for $\alpha\neq1$, it behaves like $w$CDM. It is also possible to set $\alpha=0$ from the beginning and achieve a late-time acceleration, as there would still be an integration constant. Then, one could argue that the IR cut-off could be $\propto \dot{H}$ alone as long as general covariance is respected. All these observations point towards more interesting unexplored features of GO cut-off and HDE in general. In this article, we will restrict ourselves to the case with CDM and dark energy and show that GOHDE will behave like matter in the past and transform itself into a constant towards the future. This indicates that the holography is applied to the total system and not to dark energy alone. Further details on this and the CKN entropy bound's original context will be explored elsewhere.

\section{GOHDE \& Background cosmology \label{sec:3}}

This section will examine the cosmological context that emerges from the Hubble parameter described previously in Eq. (\ref{eq:HubbleFlow}). To gain a thorough understanding, we will introduce a set of parameters, including the deceleration parameter, the equation of state for dark energy, and higher-order parameters such as jerk and snap. We will switch between redshift ($z$) and scale factor ($a$) to ensure clarity and cater to specific requirements.

\subsection*{Curious case of $\Lambda$CDM from GOHDE}

Before we derive various parameters, let's take a moment to explore a particularly intriguing case of Eq. (\ref{eq:HubbleFlow}). By setting $\alpha=1$ and $\beta=2/3$, Eq. (\ref{eq:HubbleFlow}) simplifies to the elegant form, $h^2=(1-\Omega_{\text{m0}})+\Omega_{\text{m0}}(1+z)^3$. This is precisely the Hubble parameter within the framework of the $\Lambda$CDM model. 

Fascinatingly, one can make an early inference about the values of $\alpha$ and $\beta$ based on Eq. (\ref{Standard_GOHDE}). In that context, when $\alpha=1$ and $\beta = 2/3$, the holographic dark energy density transforms into the second Friedmann equation, incorporating a cosmological constant term and CDM. This observation highlights the seemingly straightforward nature of using dark energy described by Eq. (\ref{Standard_GOHDE}) to account for the late-time acceleration in the universe. Thus, by bringing $\dot{H}$ into the picture, one brings the second Friedmann equation or the effect of pressure. 

In summary, we find ourselves at a puzzling juncture. Although a special case, if by definition Eq. (\ref{Standard_GOHDE}) corresponds the second Friedmann equation with a cosmological constant and CDM, how can it solve the fine-tuning and coincidence problems? This scenario poses a direct challenge to the principles of HDE models. Moreover, it's essential to note that this particular observation holds valid only in cases where spatial curvature or radiation density is not considered. 

A significant critique of HDE models has been their apparent \textit{inability} to recover the $\Lambda$CDM model. Here, we see that recovering it raises questions about the very foundations of HDE. The only reason it recovers the $\Lambda$ is due to the integration constant, which does not appear at the level of the field equation. At least the old cosmological constant problem is not about bringing a constant into the picture but about why this constant doesn't match the QFT vacuum energy density. 

Recent developments offer other possibilities for recovering the $\Lambda$CDM model. In our previous work \cite{Manoharan2023} and independently in the work of Moradpour et al. \cite{moradpour2023thermodynamics}, a more generic holographic formalism based on horizon thermodynamics has been studied. A more realistic approach based on minimal length was also illustrated by Luongo in \cite{Luongo2017}. These approaches adhere to holographic principles and demonstrate consistency with the laws of thermodynamics, potentially leading to new insights in this field. Simply put, one can obtain a cosmological constant as an integration constant from the thermodynamic construction of gravity \cite{PhysRevLett.75.1260}. However, this does not \textit{predict} any value for it.

In what follows, we will explore the features of various cosmic variables within the framework of GOHDE. 

\subsubsection{\textbf{Dark energy density parameter: $\Omega_\Lambda(z)$}}

To begin, we will compute the dark energy density parameter, denoted as $\Omega_\Lambda$\footnote{Here, $\Omega_i\sim\rho_i/H_0^2$ by construction, not $\rho/H^2$. We shall redefine it later.}. Given $h^2$ and the condition that $h^2=\Omega_{m}+\Omega_\Lambda$  we can readily determine $\Omega_\Lambda$ by isolating the contribution of the matter component's evolution. Therefore, we have 
\begin{align}
\Omega_\Lambda(z) = &~e^{-\frac{2(\alpha -1) x}{\beta }}\left[1 -\frac{2\Omega_{\text{m0}}}{2\alpha -3\beta-2}\left(e^{\frac{x (2 \alpha -3\beta -2)}{\beta }}-1\right)\right]\label{eq:GOHDE}\\&-\Omega_{m0} (z+1)^3\nonumber.
\end{align}
One notable aspect of this construction is that, as we approach $z\rightarrow0$, regardless of the specific values of $\alpha$ and $\beta$, we find that $\Omega_\Lambda(0)=1-\Omega_{m0}$. This means the model exhibits behaviour akin to $\Lambda$CDM in the present era. Still, depending on the values of the free parameters, it opens the door to a wide range of cosmic evolutions. Now, the most captivating question is, are $\alpha$, $\beta$, and $\Omega_{m0}$ truly independent free parameters? We'll explore this question shortly, but first, let's derive expressions for other cosmic variables. 

\subsubsection{\textbf{Dark energy equation of state parameter: $w(z)$}}
One of the most pivotal parameters under scrutiny in exploring cosmological models is the dark energy equation of state ($w$). The value of this parameter carries immense significance, as it has the potential to either validate or challenge the entire model. Numerous dedicated surveys have constrained the possible values of $w$. In the case of the $\Lambda$CDM model, $w$ is defined as -1 by default. Recent analyses, such as the Pantheon +, have placed constraints on $w$ close to $-0.90\pm0.14$ and $-0.978^{+0.024}_{-0.031}$ when incorporating the SH0ES data \cite{Brout_2022, Riess_2022}. It's important to note that while this parameter may explain late-time cosmic evolution, its ability to account for early-phase observations, such as the power spectrum, is equally vital. Therefore, setting $w$ close to -1, even in the present epoch, might not suffice to establish the credibility of a cosmological model.

Here, the GO dark energy density is a function of $z$. This aspect allows us to parametrize the model in terms of an energy density with a varying equation of state parameter. In general, for non-interacting fluids, we can write,
\begin{equation}
\Omega_\Lambda(z)=\Omega_\Lambda(z=0)\times\exp\left[3\int_{0}^{z}\frac{1+w(z')}{1+z'}dz'\right]. 
\end{equation}
The above expression holds for all forms of non-interacting energy density. Based on our prior knowledge, we set the equation of state parameters to $0$ and $1/3$ for matter and radiation, respectively. From the above relation, we can deduce that the dark energy equation of state parameter is a function of redshift and is given as
\begin{equation}
w(z)= -1+\left(\frac{1+z}{3}\right)\partial_z\ln\Omega_\Lambda(z).
\end{equation}
Here, $\partial_z$ is the derivative with respect to the redshift $z$. Now, substituting Eq. (\ref{eq:GOHDE}) into the above relation, we get, 
\begin{equation}
w(z)=\frac{-2 \alpha +3 \beta +2}{3 \beta  \left[\frac{\Omega_{\text{m0}}  \left(3 \beta
		-2 \alpha \right)}{-2 \alpha +3 \beta -2 \Omega_{\text{m0}}  +2}\left(\frac{1}{z+1}\right)^{\frac{2 (\alpha
			-1)}{\beta }-3}-1\right]}.
\end{equation}
Clearly, the equation of state of dark energy is a function of $z$ and is contingent upon the values of $\alpha$, $\beta$, and $\Omega_{\text{m0}}$. As indicated above, when the model reduces to $\Lambda$CDM, this expression simplifies to -1, ensuring that the concordance cosmology is consistently recovered.

Here lies a vital decision point. Setting $\alpha$, $\beta$, and $\Omega_{\text{m0}}$ as free parameters fixes the behaviour of $w(z)$. However, it's worth noting that $w(z)$ carries more physical significance than, for instance, $\alpha$. To accommodate this, we find a free parameter from $w(z)$ at the expense of losing control over $\alpha$. A crucial aspect to consider is that $\alpha$ remains constant throughout cosmic evolution, while $w(z)$ does not. To reconcile this, we logically choose the present value of the equation of state, denoted as $w(z=0)=w_{z_0}$, as the free parameter. Consequently, instead of $\alpha$, we introduce a new free parameter, $w_{z_0}$, which serves as a free variable. It's worth noting that $w_{z_0}=-1$ can be considered a consistency factor, as pointed out in \cite{MYUNG2007247}. In the parameter estimation sections of this manuscript, we will refer to two distinct cases. One is the GOHDE with $w_{z_0}=-1$ and the other is  $w$GOHDE, where $w_{z_0}$ is a free parameter.

With $w(z=0)=w_{z_0}$ as the present value, we have the relation,
\begin{equation}
w_{z_0}=-\frac{-2 \alpha +3 \beta -2 \Omega_{\text{m0}}  +2}{3 \beta -3 \beta \Omega_{\text{m0}}}.
\end{equation}
Given our assumption that the free parameters are constant throughout cosmic evolution, it becomes possible to determine the value of $\alpha$ based on the present value of $w$ and the other free parameters. Consequently, when we solve the expression above for $\alpha$, we obtain,
\begin{equation}
\alpha=-\frac{3}{2} \beta  \left[w_{z_0} (\Omega_{\text{m0}} -1)-1\right]-\Omega_{\text{m0}} +1.
\label{eq:alphawbO}
\end{equation}
The expression presented here aligns with findings reported in \cite{GRANDA2008275}, where they determined the equation of state by estimating pressure with the continuity equation. Here, we incorporated the boundary condition ($h^2(z=0)=1$) while solving for $h^2$, which allows us to streamline the process. When $w_{z_0}=-1$ we get, 
\begin{equation}
\alpha=\left(\frac{3}{2}\beta  -1\right)\Omega_{\text{m0}} +1.
\end{equation}
Here, when $\beta=2/3$, $\alpha$ is automatically set to unity, thus being consistent with the $\Lambda$CDM model. One crucial observation is the presence of $\Omega_{\text{m0}}$ in these expressions, which convinces us that the dark energy depends on what all are in the picture. 

With the substitution of $\alpha$ in terms of $w_{z_0}$, we will proceed to express subsequent relations in terms of $w_{z_0}$, $\beta$, and $\Omega_{\text{m0}}$. While occasionally, for brevity, we may still employ the notation $\alpha$, it's important to clarify that Eq. (\ref{eq:alphawbO}) is implicitly understood unless stated otherwise. In terms of $w_{z_0}$, the expression for $w(z)$ can be represented as,
\begin{equation}
w(z)=\frac{w_{z_0} \left[3 \beta  w_{z_0} (\Omega_{\text{m0}} -1)+2 \Omega_{\text{m0}} \right]
	\left(\frac{1}{z+1}\right)^{\frac{2 \Omega_{\text{m0}} }{\beta }+3
		w_{z_0} (\Omega_{\text{m0}} -1)}}{-3 \beta  w_{z_0}
	\left[\left(\frac{1}{z+1}\right)^{\frac{2 \Omega_{\text{m0}} }{\beta
		}+3 w_{z_0} (\Omega_{\text{m0}} -1)}-\Omega_{\text{m0}} \right]+2 \Omega_{\text{m0}} }.
\label{eq:DEoSz}
\end{equation}
Further, with $w_{z_0}=-1$ we get,
\begin{equation}
w(z)=\frac{\left[2 \Omega_{\text{m0}}-3 \beta  (\Omega_{\text{m0}} -1)\right]
	\left(\frac{1}{z+1}\right)^{\left(\frac{2}{\beta
		}-3\right) \Omega_{\text{m0}} +3}}{3 \beta  \left[\Omega_{\text{m0}}
	-\left(\frac{1}{z+1}\right)^{\frac{2 \Omega_{\text{m0}} }{\beta }-3
		\Omega_{\text{m0}} +3}\right]-2 \Omega_{\text{m0}} }.
\end{equation}

The outcomes presented above hold an intriguing aspect. Previously, we questioned the possibility of maintaining $w(z)$ as a constant. However, what we observe now is that, contingent on the values of $\beta$ and $\Omega_{\text{m0}}$, $w(z)$ exhibits a dynamic evolution. For $\beta=2/3$, we naturally arrive at $w(z)=-1$, mirroring the behaviour of the $\Lambda$CDM model. The same outcome emerges when $\Omega_{\text{m0}}$ tends toward zero, representing the de Sitter solution. If we insist on the condition $w(z)=-1$ persisting throughout cosmic history, with $\beta\neq2/3$, it imposes a specific relationship between $\beta$ and $\Omega_{\text{m0}}$, given as,
\begin{equation}
\Omega_{\text{m0}} \rightarrow \frac{3 \beta }{3 \beta -2}.
\end{equation}
An immediate inference drawn from the expression above is that when $\beta<2/3$, $\Omega_{\text{m0}}$ turns out to be negative and for $\beta>2/3$  we have $\Omega_{\text{m0}}>1$. Thus setting $w_{z_0}=-1$ for $\beta\neq2/3$ might be very tricky. 

Another notable consideration is the occurrence of singularities in the behaviour of $w(z)$ under particular parameter combinations. We will thoroughly investigate and discuss these singularities as we explore the parameter space in the next section. It's important to note that while the dark energy equation of state might exhibit singular behaviour, this doesn't necessarily imply that the effective equation of state is singular \cite{PhysRevD.106.063509}. Here, this singularity corresponds to the transition between positive and negative values of dark energy density.

\subsubsection{\textbf{Deceleration parameter: $q(z)$}}
The exploration of dark energy and the phenomenon of accelerated cosmic expansion emerged with the ground breaking measurements of the deceleration parameter, denoted as $q(z)$, as reported in \cite{Riess_1998, Perlmutter1998, Perlmutter_1997, Perlmutter_1999}. Given our initial expectations of a decelerating universe, the jargon may seem counter-intuitive for an accelerated universe. Nevertheless, $q(z)$ has become a staple metric for testing cosmological models. By making model-independent estimations of the Hubble parameter $H$ and its derivatives, we can derive several model-independent descriptors (deceleration, jerk, snap, etc.) to contrast cosmic behaviours using cosmography \cite{refId0Demianski}. The only assumption underlying the estimation of these parameters is the FLRW metric. The deceleration parameter is defined as
\begin{equation}
q(z)=-1-\frac{\dot{H}}{H^2}=-1+\frac{1+z}{h}\partial_z h.
\end{equation}

With the Hubble parameter provided by Eq. (\ref{eq:HubbleFlow}) and the relation for $\alpha$ from Eq. (\ref{eq:alphawbO}) at our disposal, we can readily express the deceleration parameter $q(z)$ in terms of $\beta$, $\Omega_{\text{m0}}$, and $w_{z_0}$ as, 
\begin{align}
&q(z)=\frac{1}{2 \beta}\Bigg\lbrace\beta -3 \beta  w_{z_0} \Omega_{\text{m0}} +3 \beta w_{z_0} -2 \Omega_{\text{m0}} \\\nonumber&\left.+\left[\frac{2 \Omega_{\text{m0}}  \left[3 \beta  w_{z_0} (\Omega_{\text{m0}}
	-1)+2 \Omega_{\text{m0}} \right]}{3 \beta  w_{z_0} (\Omega_{\text{m0}} -1)
	\left(\frac{1}{z+1}\right)^{\frac{2 \Omega_{\text{m0}} }{\beta }+3
		w_{z_0} (\Omega_{\text{m0}} -1)}+2 \Omega_{\text{m0}} }\right]\right\rbrace.
\label{eq:decelration}
\end{align}
For $w_{z_0}=-1$ we get a simpler relation that goes like, 
\begin{align}
q(z)=-1-&\frac{\Omega_{\text{m0}} }{\beta }+\frac{3 \Omega_{\text{m0}} }{2}\\\nonumber+&\frac{\Omega_{\text{m0}}  \left[2
	\Omega_{\text{m0}} -3 \beta  (\Omega_{\text{m0}} -1)\right]}{\beta  \left[2 \Omega_{\text{m0}} -3
	\beta  (\Omega_{\text{m0}} -1)
	\left(\frac{1}{z+1}\right)^{\left(\frac{2}{\beta
		}-3\right) \Omega_{\text{m0}} +3}\right]}.
\end{align}
In summary, when considering $w_{z_0}=-1$, unveil several noteworthy characteristics. Firstly, at the present epoch ($z=0$), the deceleration parameter assumes a value of $q(0)=-1+(3\Omega_{\text{m0}}/2)$. This observation places observational constraints on the current matter density. As we venture into the past ($z\rightarrow\infty$), we find that $q$ approaches 1/2 only under the condition that $\left(2-3\beta\right)\Omega_{\text{m0}}>-3\beta$. This constraint implies that exhibiting a completely matter-dominated era in the past is not obligatory. Here, dark energy itself can act like matter and give rise to the deceleration observed in the past. Consequently, this imposes constraints on $\beta$, preventing it from being arbitrarily negative. Lastly, as we approach the future, $z\to-1$, we have $q\to -1+\left[(3/2)-(1/\beta)\right]$, when $\beta(-3\beta-2\Omega_{\text{m0}}+3\beta\Omega_{\text{m0}})<0$. Therefore, a purely de Sitter phase can only manifest when $\beta=2/3$ with $w_{z_0}=-1$.

\subsubsection{\textbf{Jerk and Snap parameter: $r(z)$ and $s(z)$}}

Another crucial set of parameters capable of effectively discriminating cosmological models comprises the jerk $r(z)$ and snap $s(z)$ parameters. These parameters find valuable application in the statefinder diagnostic, which allows us to construct parametric plots capable of distinguishing a model from $\Lambda$CDM. In the case of the $\Lambda$CDM model, these parameters assume specific values, namely $\{r,s\}=\{1,0\}$, as detailed in \cite{Sahni2003}. Any deviation from these values, leading to different quadrants, enables us to classify the model. The expression for $r$ and $s$ in terms of $h$ are,
\begin{align}
r=& \frac{1}{2h^2}\frac{d^2h^2}{dx^2}+\frac{3}{2h^2}\frac{dh^2}{dx}+1,
s=-\frac{\frac{1}{2h^2}\frac{d^2h^2}{dx^2}+\frac{3}{2h^2}\frac{dh^2}{dx}}{\frac{3}{2h^2}\frac{dh^2}{dx}+\frac{9}{2}}.
\end{align}
Where, $x=\ln(a)$ and $a=1/(1+z)$ is the scale factor. 

In the context of the model we're investigating, we express $\{r,s\}$ as functions of $x=\ln(a)$, where $\alpha$, $\beta$, and $\Omega_{\text{m0}}$ serve as free parameters. The corresponding equations are,
\begin{equation}
r(x) = \frac{(\alpha -1) (2 \alpha -3 \beta -2) (2 \alpha -3 \beta
	+2 \Omega_{\text{m0}} -2)}{\beta ^2 \left[2 \alpha -3 \beta -2 \Omega_{\text{m0}} 
	e^{\frac{x (2 \alpha -3 \beta -2)}{\beta }}+2 \Omega_{\text{m0}}
	-2\right]}+1,
\label{eq:jerk}
\end{equation}
and 
\begin{equation}
s(x) = \frac{2 (\alpha -1)}{3 \beta }.
\label{eq:snap}
\end{equation}
While evaluating the GOHDE model, we can fix the value of $\alpha$ based on Eq. (\ref{eq:alphawbO}). What is particularly intriguing is that the snap parameter remains constant when both $\alpha$ and $\beta$ are constant. This implies that unless $\alpha$ approaches one or $\beta$ approaches infinity, the model never exhibits behaviour identical to that of $\Lambda$CDM. Even in the scenario where $\beta\to2/3$, achieving a true $\Lambda$CDM behaviour necessitates the condition $w_{z_0}\to-1$. In most dark energy models, we observe a future or past $\Lambda$CDM nature. Unfortunately, this characteristic is notably absent in the construction of GOHDE. A constant snap parameter indicates a linear relation between jerk and deceleration parameters, a non-trivial observation. Furthermore, as $z\to -1$, we find that $r$ approaches $(2-2\alpha+\beta)(1-\alpha+\beta)/\beta^2$. It's worth noting that these observations come with the stipulations that $2\alpha<3\beta$ and $\beta>0$. One could argue that $\Lambda$CDM can be viewed as a unique version of GOHDE, not any asymptotic limit.

\subsubsection{\textbf{Horizon entropies: $S_H(z)$ and $S_L(z)$}}

Exploring the thermodynamic aspects of cosmological models has gained significant prominence, spurred by the recognition of Einstein's equation as a thermodynamic equation of state with cosmological constant as an integration constant \cite{PhysRevLett.75.1260}. Additionally, the foundation of the HDE concept rests upon the notion of a UV-IR relation constrained by the horizon entropy. Consequently, the horizon entropy is one of the most pivotal and indispensable parameters associated with the HDE models. The choice of entropy is unambiguous in the standard HDE framework, employing the Hubble cutoff and the Bekenstein-Hawking area law. In this scenario, the entropy of central interest is that of the Hubble horizon\footnote{For a non-flat universe, the apparent horizon becomes the relevant horizon for accounting for horizon entropy, although it reduces to the Hubble horizon when $k=0$.}. Within the $\Lambda$CDM model, the Hubble horizon exhibits a characteristic entropy maximization toward the future, aligning with the principles of the generalized second law of thermodynamics. This entropy maximization signifies that the entropy reaches its maximum limit as $z\to -1$, a feature primarily driven by the presence of the cosmological constant \cite{PhysRevD.96.063513}.

There are two ways to HDE density,
\begin{equation}
\rho_{\Lambda}\sim S/L^4 \text{ and } \rho_{\Lambda}\sim L^{-2}.
\end{equation}
Both these are equivalent in the standard picture, where $S\sim L^2$. In this article, we shall not consider the first version where we can modify both $S$ and $L$ as in \cite{Dheepika2022a, doi:10.1142/S0218271822501073}. However, one feature that remains identical in both these constructions is that $S$ is a function of the chosen $L$.

Now, it might seem less intuitive to investigate the entropy of the Hubble horizon, as the Hubble cut-off is not always the IR cut-off. An argument in favour of assigning horizon entropy to the Hubble horizon is that the expression for the Hubble flow inherently encapsulates the effects of the choice of IR cut-off. However, this appears unclear, as we start with $S\sim L^2$ and then investigate $S\sim H^{-2}$ for cases with $L\neq 1/H$. Although there is nothing wrong with studying both, we must see whether we started from some ill-defined entropy if $S\sim L^2$.

Here, the first entropy corresponds to the one associated with the Hubble horizon, denoted as $S_H(z)$, while the second represents the entropy related to the IR cut-off $(L=L_{GO})$, indicated as $S_L(z)=S_{GO}(z)$. This dual exploration allows us to better understand the thermodynamic properties within the HDE framework. Here we have the expressions for $S_H(z)$ and $S_{GO}(z)$ given as,
\begin{widetext}
	\begin{align}
	S_H(z) =& \frac{3 \beta  w_{z_0} (\Omega_{\text{m0}} -1)+2 \Omega_{\text{m0}} }{(z+1)^3 \left[3 \beta  w_{z_0} (\Omega_{\text{m0}} -1) \left(\frac{1}{z+1}\right)^{\frac{2 \Omega_{\text{m0}} }{\beta
			}+3 w_{z_0} (\Omega_{\text{m0}} -1)}+2 \Omega_{\text{m0}} \right]},\label{eq:BHS}\\
	S_{GO}(z) =& -\frac{3 \beta  w_{z_0} (\Omega_{\text{m0}} -1)+2 \Omega_{\text{m0}} }{(\Omega_{\text{m0}} -1) (z+1)^3 \left\lbrace 2 \Omega_{\text{m0}} -3 \beta  w_{z_0} \left[\left(\frac{1}{z+1}\right)^{\frac{2
				\Omega_{\text{m0}}}{\beta }+3 w_{z_0} (\Omega_{\text{m0}} -1)}-\Omega_{\text{m0}} \right]\right\rbrace }.\label{eq:GOS}
	\end{align}
\end{widetext}
The concept of entropy maximization's validity hinges on the specific values of the free parameters involved. One can conduct tests to assess this by calculating the first and second derivatives of the respective entropy expressions for various combinations of parameter values. Alternatively, plotting these entropies can draw a quick inference.

To avoid huge values on the scale, we examine the entropies in the following manner. Traditionally, the horizon entropy is defined as,
\begin{equation}
S(L)=\frac{\pi L^2 k_B}{\ell_p^2}.
\end{equation}
Here, $\ell_p^2=G\hbar/c^3$. When considering the Hubble cut-off, one could assume $L=c/H$; for other generic cut-offs, we have $L=c/L_{IR}$. Either way, we have
\begin{equation}
S(L)=L^2\left(\frac{\pi k_Bc^5}{G\hbar}\right).
\end{equation} 
However, this results in a huge number on the order of $10^{120}$ in units of $k_B$. Instead of directly computing $S(L)$, we compute $S(L)\times H_0^2$ and express them in units of ${\pi k_Bc^5}/{G\hbar}$. Consequently, the present value of $S$ with the Hubble cut-off is normalized to unity. This procedure scales the numerical value and does not affect the inherent characteristics of $S$.

Now that we have introduced the cosmic variables of interest, our next objective is to examine their characteristics within the parameter space. It's worth emphasizing that the forthcoming analysis focuses on illuminating each variable's unique attributes in the parameter space. In the subsequent section, we will determine the best-fit values for these free parameters.

\section{Behaviour of Cosmological variables under GOHDE \label{sec:4}}

This section will verify whether the GOHDE model under scrutiny can consistently align with observations and theoretical concepts. We will explore the parameter space defined by $\beta$ and $w_{z_0}$ to achieve this goal. We will assess whether the cosmic variables described earlier exhibit reasonable evolutionary behaviour.

It's important to note that, for this analysis, we will adopt specific values: a present Hubble parameter of $H_0=70~\text{km/s/Mpc}$ and a current matter density of $\Omega_{\text{m0}}=0.3$. However, it's crucial to emphasize that these values are chosen solely for illustrative purposes. In the subsequent section, we will refine these estimates based on observational data. Throughout our analysis, we will consider a comprehensive range of parameter values, encompassing all possible combinations of $\beta\in(0.6, 2/3, 0.7)$ and $w_{z_0}\in(-0.9, -1.0, -1.2)$. While this selection may initially appear biased, it serves its purpose. By including these values, we cover the well-established $\Lambda$CDM case and potential deviations from it. Due to their proximity, we anticipate these chosen values may exhibit similar characteristics in certain regions, particularly in the present epoch. However, it's essential to acknowledge that they might reveal unusual behaviours in more distant cosmic epochs. The presence or absence of such features will play a crucial role in determining the model's credibility.

\subsubsection{\textbf{A quick illustration test}\label{subsec:quick}}
The most straightforward approach to assess the model's alignment with observational data is to superimpose the data points onto the corresponding cosmological model's predicted evolution. In this context, we plot the Hubble flow as defined by Eq. (\ref{eq:HubbleFlow}) with the Observational Hubble Data \cite{SUDHARANI2023250}. Additionally, we display the apparent magnitude progression of Type Ia Supernovae alongside the Pantheon Sample data \cite{Scolnic_2018}.
\begin{figure}[h]
	\includegraphics[width=0.48\textwidth]{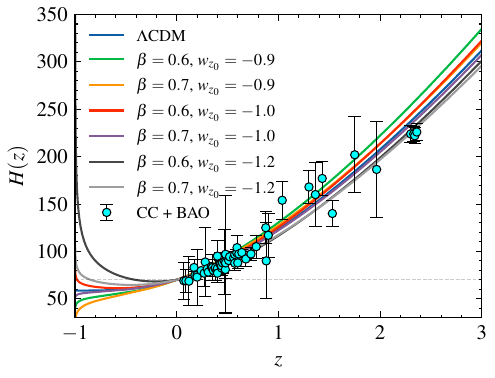}
	\caption{(colour online) The graph illustrates $H(z)$ as a function of redshift $(z)$ for various combinations of $\beta$ and $w_{z_0}$. The data points represent the Observational Hubble Data (OHD) sourced from \cite{SUDHARANI2023250}.}
	\label{fig:OHD_data}
\end{figure}

\begin{figure}[h]
	\includegraphics[width=0.48\textwidth]{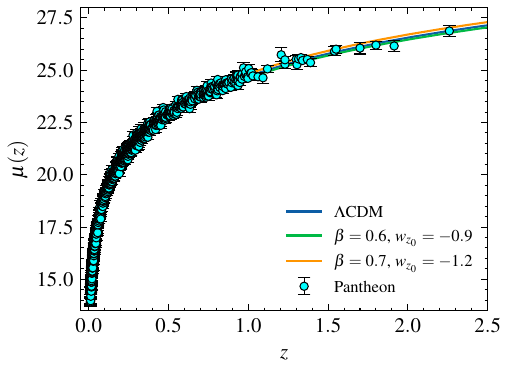}\\
	\caption{(colour online) The graph illustrates the apparent magnitude of type Ia Supernovae $\mu(z)$ as a function of redshift $(z)$ for various combinations of $\beta$ and $w_{z_0}$. The data points represent the Pantheon sample sourced from \cite{Scolnic_2018}. The figure suggest that, variations in the free parameter is hard to acknowledge.}
	\label{fig:PAN_data}
\end{figure}

In Fig. (\ref{fig:OHD_data}), all the curves exhibit similar behaviour up to $z=0$ and tend to approach Phantom, quintessence, and de Sitter-like behaviour as $z$ approaches $-1$. Among the various parameter combinations, the de Sitter behaviour is exclusively observed in the curve corresponding to $(\beta, w_{z_0}) = (2/3, -1)$, which is the $\Lambda$CDM scenario. When $\beta$ deviates from 2/3, a lower value of $\beta$ with $w_{z_0} = -1$ displays a phantom-like nature, whereas a higher value behaves like quintessence dark energy in the future. However, this correlation with $\beta$ disappears once $w_{z_0}$ diverges from the phantom divide. If $w_{z_0}$ is positioned below the Phantom divide, it consistently exhibits a Phantom behaviour, regardless of the value of $\beta$. Conversely, when $w_{z_0}$ lies above the Phantom divide, it acts like a quintessence field. Given that de Sitter is renowned for its stability in various contexts \cite{PhysRevD.84.044040, PhysRevLett.89.081301}, any departure from the $\beta = 2/3$ value may signify the presence of new physics.

Another testing ground for cosmological models involves Type Ia Supernovae (SNe Ia) measurements. These supernovae serve as standardized cosmic rulers, allowing us to gauge distance scales accurately. The Pantheon dataset represents one of the most extensive compilations (1048 data points) of apparent magnitudes for SNe Ia across various redshifts \cite{Scolnic_2018}. Using the background cosmology, we can estimate the apparent magnitude using Eqs. (\ref{eq:apparentmagnitude}) and (\ref{eq:D_L}). We plot the Pantheon data with the corresponding expressions mentioned above in Fig. (\ref{fig:PAN_data}). Distinguishing differences between the various curves is challenging, and the chosen combinations of $\beta$ and $w_{z_0}$ explain the observations. Statistical tests must be conducted to identify the appropriate values of the free parameters.

\subsubsection{\textbf{Behaviour of cosmic variables}}
Parameters close to the concordance value effectively account for late-time cosmic acceleration. To explore deeper into cosmic behaviour, we will systematically examine the characteristics of each cosmic variable. For illustration, we present most cosmic parameters plotted against $1+z$ on a logarithmic scale. This approach facilitates the identification and interpretation of asymptotic behaviours in the universe's past and future.

\begin{figure}[h]
	\includegraphics[width=0.48\textwidth]{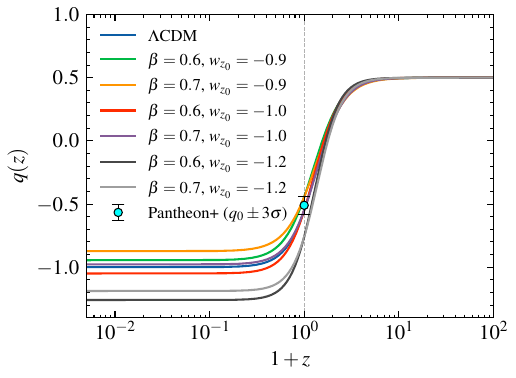}
	\caption{(colour online) The diagram illustrates the deceleration parameter $q(z)$ as a function of redshift $(z)$ plotted against $1+z$ in log scale for different values of $\beta$ and $w_{z_0}$. The data point corresponds to the Pantheon+ best estimate with the error bar enlarged to 3$\sigma$ \cite{Brout_2022}.}
	\label{fig:q_z}
\end{figure}

Figure (\ref{fig:q_z}) illustrates the evolution of the deceleration parameter, given by Eq.(\ref{eq:decelration}), as derived in the previous section.
Indeed, all combinations of parameters lead to late time acceleration, and the current value of the deceleration parameter aligns well with various choices of $\beta$ and $w_{z_0}$. Distinguishing between phantom and quintessence behaviour is also straightforward by examining the $q(z)$ value as $1+z$ approaches zero. We can also see that $w_{z_0}=-1.2$ is disfavoured above $3\sigma$, putting a solid constraint over the values of free parameters.

\begin{figure}[t]
	\includegraphics[width=0.48\textwidth]{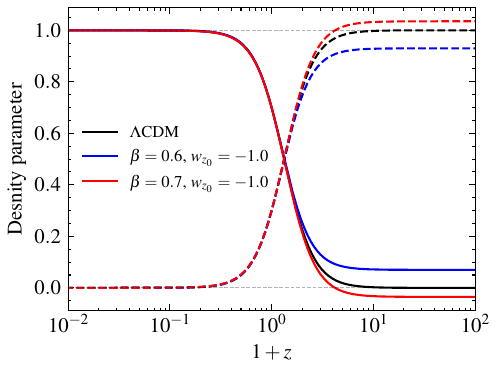}
	\caption{(colour online) Density parameter $\tilde{\Omega}(z)$ as a function redshift $(z)$ plotted against $1+z$ in log scale for different values of $\beta$ with $w_{z_0}=-1$. The solid line corresponds to $\tilde{\Omega}_{\Lambda}$ and dashed line to $\tilde{\Omega}_{m}$. Each colour represents different combinations of free parameters.}
	\label{fig:DenPar1}
\end{figure}

The central question we aimed to explore is whether a model's capability to explain late time acceleration ensures its consistency throughout cosmic evolution. Here, we first study the characteristics of density parameters during cosmic evolution. For understanding of the characteristics of dark energy density, it is convenient to work with $\rho_{\Lambda}/H^2$ rather than $\rho_{\Lambda}/H_0^2$, which we derived earlier to find the equation of state. To distinguish between these definitions, we introduce the normalized density parameter, denoted as $\tilde{\Omega}_i=\Omega_i/h^2$, where the sum of all $\tilde{\Omega}_i$ equals 1 for a flat universe. An unusual feature can be spotted in Fig. (\ref{fig:DenPar1}). Here, a genuine matter-dominated era is only achievable within the $\Lambda$CDM framework or when $\beta=2/3$ with $w_{z_0}$ fixed at $-1$. While the GOHDE model explains the late-time dark energy dominance, it fails to reproduce a proper matter-dominated past. Deviations, as observed in the context of GOHDE, lead to the dark energy density settling at either a negative or positive value, depending on the values of $\beta$ with $w_{z_0}$ fixed at $-1$. Although the constraint $\sum\tilde{\Omega}_i=1$ is maintained, such values often conflict with other observations, such as the CMB power spectrum, baryon density, etc. In the computations related to the Cosmic Microwave Background (CMB), particularly concerning density perturbations, we typically assume a smooth dark energy or a dark energy equation of state close to $-1$. However, this assumption is false when a genuine matter-dominated phase is absent with dark energy that scales like matter, where we need to consider dark energy perturbations. Otherwise, this must be some decay model with interactions between the dark sectors. To address this, one can conduct a surface-level examination within the framework of the relative smoothness condition, as outlined in \cite{PhysRevD.78.087303} using the CAMB module \cite{cambpython}. Note that while this approach provides a helpful illustration, it may not always accurately capture the complexity of the situation. Such deeper investigations are beyond the scope of this manuscript and are kept for future work. 

\begin{figure*}
	\includegraphics[width=0.48\textwidth]{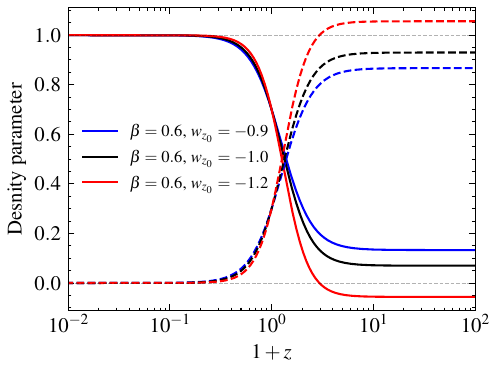}
	\includegraphics[width=0.48\textwidth,height=0.365\textwidth]{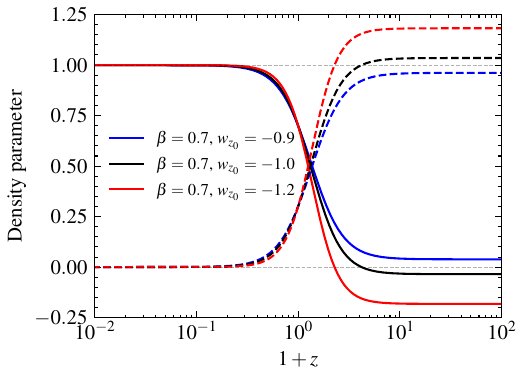}
	\caption{(colour online) Density parameter $\tilde{\Omega}(z)$ as a function redshift $(z)$ plotted against $1+z$ in log scale for different values of $\beta$ and $w_{z_0}$. The solid line corresponds to $\tilde{\Omega}_{\Lambda}$ and dashed line to $\tilde{\Omega}_{m}$. Each colour represents different combinations of free parameters.}
	\label{fig:DenPar2}
\end{figure*}

\begin{figure}[h]
	\includegraphics[width=0.48\textwidth]{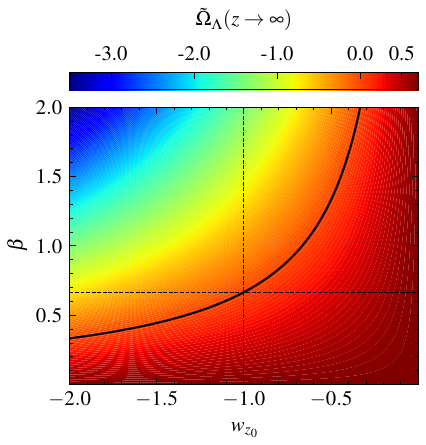}
	\caption{(colour online) Value of $\tilde{\Omega}_{\Lambda}(z\to\infty)$ for different values of $\beta$ and $w_{z_0}$. The solid curve corresponds to $\tilde{\Omega}_{\Lambda}(z\to\infty)=0$ case given by $\beta=-2/(3w_{z_0})$.}
	\label{fig:MatterDomi}
\end{figure}

Figure (\ref{fig:DenPar2}) illustrates the cases for $w_{z_0}\in(-0.9,-1.0,-1.2)$ with $\beta\in(0.6,0.7)$. Clearly, $\tilde{\Omega}_{\Lambda}$ settle at a non-zero value, and is given by,
\begin{equation}
\tilde{\Omega}_{\Lambda}(z\to\infty)=-\frac{1}{2}(2+3w_{z_0}\beta)(\Omega_{m0}-1).
\end{equation}
The above expression tends to zero only when $\beta\to 2/3$ for $w_{z_0}\to-1$, else it must satisfy the relation $\beta=-2/(3w_{z_0})$ for non zero $\Omega_{m0}$ (See Fig. (\ref{fig:MatterDomi}) with $\Omega_{\text{m0}}=0.3$). In a more general setting, for non-zero $\beta$ we have $\tilde{\Omega}_{\Lambda}(z\to\infty)\to 0$, provided $\beta  \left[3 \beta  w_{z_0} (\Omega_{\text{m0}} -1)+2 \Omega_{\text{m0}} \right]>0$, $\left\lbrace3 \beta  \left[w_{z_0} (\Omega_{\text{m0}} -1)-1\right]+2 \Omega_{\text{m0}} \right\rbrace \left[3 \beta 
w_{z_0} (\Omega_{\text{m0}} -1)+2 \Omega_{\text{m0}} \right]<0$ and $3 \beta ^2 w_{z_0} (3 w_{z_0}+1)+\Omega_{\text{m0}}^2 (3 \beta 
w_{z_0}+2)^2>\beta\Omega_{\text{m0}} (6 w_{z_0}+1)  (3 \beta  w_{z_0}+2)$. As an additional note, when we refer to $z\to\infty$, we indicate the range where we anticipate a significant matter-dominated phase rather than a radiation-dominated era. This distinction arises from our initial omission of radiation density. Practically, $z\to\infty$ implies a very large $z\gg0$. 

Let's see the reasons behind this phenomenon. This behaviour arises directly from the inherent properties of dark energy. GOHDE's dependence on the matter component's characteristics is critical to this event. As depicted by Eq. (\ref{eq:GOHDE}), the evolution of $\Omega_{\Lambda}$ is intimately connected to the dynamics of $\Omega_m$, which scales as $(1+z)^3$. Consequently, unless this correlation is effectively countered, dark energy exhibits behaviours similar to diffuse matter, with its equation of state gradually approaching zero. This condition significantly influences the equations governing dark energy perturbations and subsequently impacts cosmic history. The only way to remove such behaviour is to find correlations between $\alpha$ and $\beta$. From the general solution, it was clear that when $\alpha=3\beta/2$, the matter-like behaviour will be removed, which explains why $\beta=2/3$ behaves exactly like $\Lambda$CDM. This, however, retains the behaviour of radiation in general. Thus, dark energy will scale like radiation in the very early phase. These observations were never reported earlier. Indirectly, hints were obvious from Ricci HDE, where $\alpha=2\beta$ removes radiation-like scaling. Sometimes, these transitions leave an imprint on $w$.

Let's now analyse the characteristics of the dark energy equation of state parameter, denoted $w(z)$. Building upon our earlier investigation of the density parameter's behaviour, we observed that dark energy exhibits traits resembling those of diffuse matter. This indicates that the effective equation of state must approach zero in the past, implying that dark energy behaves indistinguishably from matter in earlier cosmic epochs.

A notable observation arises from this matter-like behaviour, which is readily apparent when examining the dark energy equation of state as given in Eq. (\ref{eq:DEoSz}). It becomes evident from this equation that for any combination of $\beta$ and $w_{z_0}$, except when they are precisely $(2/3,-1)$, under the conditions $\beta  \left[3 \beta  w_{z_0} (\Omega_{\text{m0}} -1)+2 \Omega_{\text{m0}} \right]>0$ and $3 \beta ^2 w_{z_0} (3 w_{z_0}+1)+\Omega_{\text{m0}}^2 (3 \beta 
w_{z_0}+2)^2>\beta  (6 w_{z_0}+1) \Omega_{\text{m0}} (3 \beta  w_{z_0}+2)$, the value of $w(z)$ approaches zero towards the past. This behaviour is odd for dark energy, implying that it behaves similarly to matter, leading to dark energy perturbations.

\begin{figure}[h]
	\includegraphics[width=0.49\textwidth]{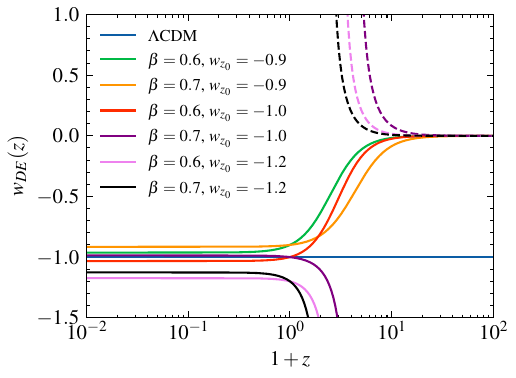}
	\caption{(colour online) The diagram illustrates dark energy equation of state, $w(z)$, as a function of redshift $(z)$ plotted against $1+z$ in log scale for different values of $\beta$ and $w_{z_0}$. As the dark energy becomes negative, the corresponding equation of state becomes complex and we plot the absolute value with dashed line.}
	\label{fig:DEoS}
\end{figure}

This observation has consequences, especially when comparing it to one of the significant drawbacks of the HDE model with the Hubble cut-off as the infrared cut-off. In that model, the equation of state remained matter-like throughout the cosmic evolution. However, in the scenario presented here, while the present value of $w(z)$ accounts for the current accelerated expansion of the universe, it tends toward zero in the past. This behaviour is precisely why the energy density discussed earlier displayed unusual characteristics. The asymptotic nature of $w(z)$ for various parameter combinations is shown in Fig. (\ref{fig:DEoS}).

Additionally, as depicted in Fig. (\ref{fig:DEoS}), we observe singular points for specific combinations of parameters. These are direct consequences of dark energy becoming negative \cite{PhysRevD.106.063509}, and we can readily pinpoint these points when,
\begin{align}
z= -1+
\exp\left\lbrace\left[{\frac{\beta }{3 \beta w_{z_0} (\Omega_{\text{m0}} -1)+2\Omega_{\text{m0}} }}\right]\right.\label{eq:singularz}\\\nonumber\left.\log\left(\frac{3\beta  w_{z_0}}{3 \beta  w_{z_0} \Omega_{\text{m0}} +2
	\Omega_{\text{m0}} }\right)\right\rbrace.
\end{align}
It's important to note that while singularities exist in the dark energy equation of state, they do not appear in the total effective equation of state. One can show this analytically by looking at the deceleration parameter. Consequently, we observe a smooth transition to the decelerated phase as we extend further into the past. If the dark energy remains positive, the equation of state goes from negative to zero without singular points. 

Another intriguing aspect to highlight is the transition between phantom and quintessence behaviours. For instance, in Fig. (\ref{fig:DEoS}), we can observe that for the parameter combination $(\beta,w_{z_0})=(0.7,-0.9)$, the value of $w(z)$ never crosses the phantom divide. It remains relatively constant towards the future and gradually tends towards zero in the past. However, for $(\beta,w_{z_0})=(0.7,-1.0)$, although it appears as quintessence dark energy in the future, it exhibits a phantom crossing in the past. It crosses the phantom divide and exhibits a singular point whose location is consistent with the expression given above. However, all Phantom crossing does not imply a singular point. For instance, when $(\beta,w_{z_0})=(0.6,-1.0)$, even though it crosses the phantom divide, it does not display such singular characteristics. Thus, phantom crossing can be classified into future and past based on the presence of singular points. 

In summary, while all these parameter combinations can explain late-time acceleration and align with local observations, they diverge significantly from the $\Lambda$CDM model when extended further into the past. The only exception is the $(\beta,w_{z_0})=(2/3,-1)$ combination.

\subsubsection{\textbf{Being different from $\Lambda$CDM}}
The most striking feature of GOHDE is the ability to recover $\Lambda$CDM as a special case. However, we have seen that the parameter space also accommodates deviations from it. Now, the question is whether these deviations resemble $\Lambda$CDM at any epoch. To investigate this, we can leverage the jerk and snap parameters. First of all, the snap parameter remains independent of redshift ($z$) and maintains a constant value as given by Eq. (\ref{eq:snap}). This value aligns with the $\Lambda$CDM scenario exclusively when $\alpha=1$, which is possible when, 
\begin{equation}
\beta=-\frac{2 \Omega_{\text{m0}} }{3 (w_{z_0} \Omega_{\text{m0}} -w_{z_0}-1)}.
\end{equation}
Across different combinations, we observe that $s$ exhibits both positive and negative values, indicative of dark energy resembling both quintessence and Chaplygin gas. This trend is also reflected in the behaviour of the jerk parameter Eq. (\ref{eq:jerk}). In contrast to the snap parameter, the jerk parameter varies with redshift $(z)$. We have already established a linear relationship between the jerk and deceleration parameters, which becomes evident due to the snap parameter's constancy. Therefore, the jerk parameter serves as a classifier, akin to $q(z)$, to distinguish between Phantom and quintessence behaviours.
\begin{figure}[h]
	\includegraphics[width=0.48\textwidth]{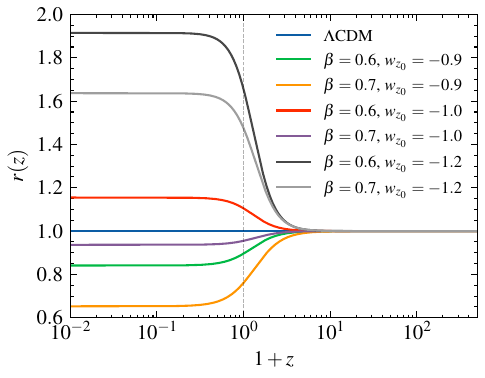}
	\caption{(colour online) The graph shows the jerk parameter $r(z)$ as a function of redshift $(z)$ plotted against $1+z$ in log scale for different values of $\beta$ and $w_{z_0}$. For $\Lambda$CDM $r(z)=1$. }
	\label{fig:jerkp}
\end{figure}

As illustrated in Fig. (\ref{fig:jerkp}), the system initially exhibits $r(z)=1$ in the past, somewhat similar to $\Lambda$CDM, before evolving to values both above and below $1$ for various combinations of $\beta$ and $w_{z_0}$. Notably, no parameter combinations yield a future behaviour resembling $\Lambda$CDM unless the model is inherently $\Lambda$CDM by definition. This observation sets this model apart from most HDE models, which often display past or future $\Lambda$CDM-like nature. It's important to note that while the jerk parameter $r$ remains constant at $1$ in the past, this doesn't necessarily imply a past identical to $\Lambda$CDM, as the snap, $s\neq 0$.

\subsubsection{\textbf{Puzzling horizon entropy}}
We encounter two options when analysing the horizon entropy, as given in Eqs. (\ref{eq:BHS}) and (\ref{eq:GOS}). Firstly, we can explore the properties of the Hubble horizon based on the Hubble parameter and apply the Bekenstein-Hawking area law. Alternatively, we can derive entropy from the dark energy expression. Both approaches represent a form of entropy, as discussed previously, and which proves more suitable will become evident in the subsequent analysis.

\begin{figure}[b]
	\includegraphics[width=0.48\textwidth]{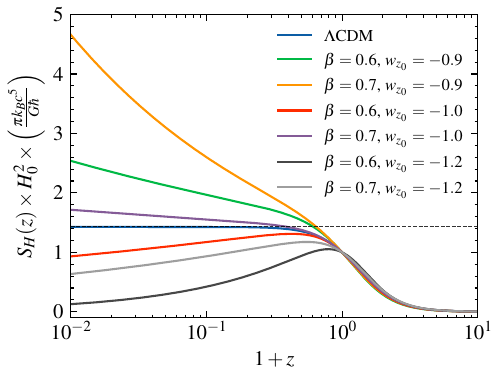}
	\includegraphics[width=0.48\textwidth]{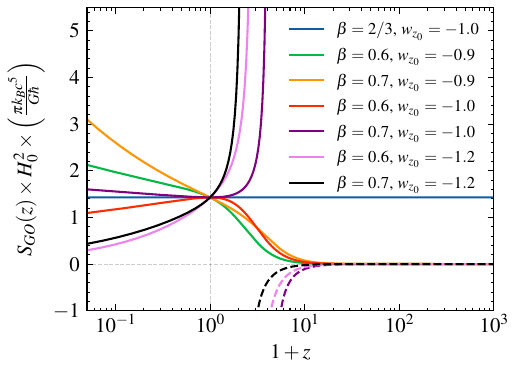}
	\caption{(colour online)Hubble horizon entropy $(S_H(z))$ and GO horizon entropy $(S_{GO}(z))$ as a function of redshift $(z)$ plotted against $1+z$ in log scale for different values of $\beta$ and $w_{z_0}$.}
	\label{fig:HEntropy}
\end{figure}

In adopting the first strategy, utilising the expression in Eq. (\ref{eq:BHS}), we observe a reasonable thermodynamic behaviour. The condition for entropy maximisation is automatically met for $\Lambda$CDM, indicating a cosmological constant. For behaviours resembling quintessence, entropy steadily increases without bounds, while for Phantom-like behaviours, it violates the second law of thermodynamics and tends towards zero. These characteristics are clearly illustrated in the first figure within Fig. (\ref{fig:HEntropy}). Furthermore, it's apparent that the entropy exhibits similar characteristics up to the present time and diverges in the future. The normalisation scheme we employed allows us to fix the current entropy value at one, and it saturates to a maximum in the end only for the $\Lambda$CDM model.
As additional details, it's essential to discuss the concept of entropy maximisation within this context. When we focus on the Hubble horizon and define entropy using an area law, we can only attain entropy maximisation in the presence of a cosmological constant. Why is this the case? Only the cosmological constant can ensure that the Hubble flow approaches a finite constant in the future. While there are models where the dark energy equation of state tends towards $-1$ in the future, it's crucial to note that such behaviour doesn't always guarantee entropy maximisation. However, there can be exceptions \cite{Manoharan2023, Li_2019}.

Let's now see the thermodynamics by choosing the GO cut-off as our horizon of interest, as defined in Eq. (\ref{eq:GOS}). This discussion presents a fundamental question: why study the thermodynamics of such a horizon? This seemingly simple question challenges the core principles of the standard HDE paradigm.

In the conventional approach, we define dark energy as $\rho_{\Lambda}\sim S/L^4$. To define $\rho_{\Lambda}\sim L^{-2}$, we must assume that $S\sim L^2$. In the construction of HDE, the entropy is defined with the chosen IR cut-off, not the Hubble horizon. Alternatively, one could consider defining HDE as $\rho_{\Lambda}\sim 1/(H^2L^4)$, but this is not the conventional way. While such an approach might be interesting as a phenomenological exploration, it's beyond the scope of our current discussion.

Returning to the definition of HDE, the entropy of the holographic screen is taken as $S\sim L^2$, where here, the GO cut-off serves as $L$. If we had chosen the Hubble scale as the cut-off, there would be no ambiguity regarding the behaviour of entropy. However, when other IR cut-offs are considered, it needs to be clarified why the Hubble scale exhibits the appropriate thermodynamic behaviour. If the Hubble scale were the only scale capable of producing such a response, it would be puzzling why $S\sim L^2$ was considered for the definition of GOHDE. 

As shown in the second figure of Fig. (\ref{fig:HEntropy}), intriguing features emerge that support the earlier arguments. Notably, for the parameter combination $(\beta,w_{z_0})=(2/3,-1.0)$, $S_{GO}$ coincides with the saturation limit of $S_{H}$ under the same conditions. This alignment is not coincidental; it reflects that $\rho_{\Lambda}$ represents the cosmological constant under these circumstances. While we can identify Phantom and quintessence-like characteristics towards the future, the past reveals singular points and instances of negative entropy. These findings raise questions and pose further challenges to the general definition of HDE. Thus, we think the standard definition of HDE may not represent a proper thermodynamic boundary connecting UV and IR when we assume scales other than $L=1/H$. 

When the universe attains a pure de Sitter state, the Hubble flow becomes a constant, and the square of the Hubble flow equates to the energy density of the cosmological constant. Consequently, entropy is the natural inverse of the energy density based on the first law of horizon thermodynamics. While this observation holds for the pure de Sitter universe and the total energy, it raises questions about why it should hold for an individual component, particularly for dark energy. Thus, HDE itself might not be a well-posed construction.

Another observation is that the GO horizon's entropy isn't properly normalised according to our scheme. Notably, for the $\Lambda$CDM case, this value already reaches the saturation limit of the Hubble horizon entropy. When entropy is saturated, unless some unknown physics is at play, there's no reason to expect the universe to undergo further evolution. Thus, the horizon considered for HDE construction isn't just a thermodynamic horizon; instead, it is the holographic screen that sets the entropy bound. This suggests that the holographic principle offers a means to define the total energy density, and it is this idea that we should extract from the seminal work of Cohen-Kaplan-Nelson \cite{PhysRevLett.82.4971}. The correspondence between the GO cut-off and the second Friedmann equation for a cosmological constant for $(\beta,w_{z_0})=(2/3,-1.0)$ lends further support to this explanation. In fact this result is consistent with the Komar energy used to construct cosmological models in emergent paradigm  \cite{PhysRevD.92.024051, HassanVT_2022}. 

\subsubsection{\textbf{Age of the Universe}}
One of the most significant derived quantities in any cosmological model is the age of the universe. The age estimation is closely related to the present value of the Hubble parameter, denoted as $H_0$, and hence holds implications for addressing other pressing issues, such as the Hubble tension \cite{Brout_2022}. Astrophysical observations have imposed strong constraints on the age \cite{Cimatti_2023}. While local measurements claim to be model-independent and the age derived from the Hubble flow is model-dependent, estimating the age serves as a testing ground for addressing the Hubble tension. Our primary focus here is not to resolve the Hubble tension but to utilize age as a derived quantity to explore the behaviour of the GOHDE model within the $(\beta, w_{z_0})$ parameter space. Considering existing tensions, we aim to investigate whether deviations from the $\Lambda$CDM model align with local observations.

The age of the universe is defined as the cosmic time elapsed from the scale factor reaching from ``zero'' to ``one''. In simpler terms, it represents the duration between the ``Big Bang'' and the present moment. For our analysis, we shall accept this definition, and the expression for the age of the universe is given by,
\begin{equation}
\text{Age} = \int_{0}^{a=1} \frac{1}{aH(a)} da.
\end{equation}

We can perform the above integral to obtain the analytical expression for the universe's age by using the equation for the Hubble parameter in Eq. (\ref{eq:HubbleFlow}). Here, the age as a function of `$a$' under GOHDE takes the form, 
\begin{widetext}
	\begin{equation}
	\text{Age}_{\text{GOHDE}}(a) = \frac{_2F_1\left(\xi_1,\xi_2;\xi_3;\xi_4\right) 2\sqrt{\frac{3 \beta  w_{z_0} (\Omega_{\text{m0}} -1) a^{\frac{2
						\Omega_{\text{m0}} }{\beta }+3 w_{z_0} (\Omega_{\text{m0}} -1)}}{2 \Omega_{\text{m0}} }+1} \,
	}{3 H_0 \sqrt{\frac{3 \beta  w_{z_0}
				(\Omega_{\text{m0}} -1) a^{\frac{2 \Omega_{\text{m0}} }{\beta }+3 w_{z_0} (\Omega_{\text{m0}}
					-1)}+2 \Omega_{\text{m0}} }{a^3 (3 \beta  w_{z_0} (\Omega_{\text{m0}} -1)+2 \Omega_{\text{m0}}
				)}}}
	\label{eq:ageGOHDE}
	\end{equation}
	where, $_2F_1\left(\xi_1,\xi_2;\xi_3;\xi_4\right)$ is the hypergeometric function \cite{822801MathFun} with,
	\begin{align*}
	\xi_1=&\frac{1}{2},~\xi_2=\frac{3 \beta }{6 w_{z_0} \beta 
		(\Omega_{\text{m0}} -1)+4 \Omega_{\text{m0}} },~
	\xi_3=1+\frac{3 \beta }{6 w_{z_0} \beta 
		(\Omega_{\text{m0}} -1)+4 \Omega_{\text{m0}} },~\xi_4=&\frac{3 a^{3 w_{z_0} (\Omega_{\text{m0}}
			-1)+\frac{2 \Omega_{\text{m0}} }{\beta }} w_{z_0} \beta  (\Omega_{\text{m0}} -1)}{-2
		\Omega_{\text{m0}} }.
	\end{align*}
	The above equation for $(\beta,w_{z_0})=(2/3,-1.0)$  reduces to,
	\begin{equation}
	\text{Age}_{\text{GOHDE}}(\beta=2/3,w_{z_0}=-1) = \frac{2 i \tan
		^{-1}\left(\frac{a^{3/2} \sqrt{1-\Omega_{\text{m0}} }}{\sqrt{a^3 (\Omega_{\text{m0}}
				-1)-\Omega_{\text{m0}} }}\right)}{3  H_0 \sqrt{1-\Omega_{\text{m0}}}}=\frac{2\sinh^{-1}\left(\sqrt{\frac{a^3(1-\Omega_{\text{m0}})}{\Omega_{\text{m0}}}}\right)}{3H_0\sqrt{1-\Omega_{\text{m0}}}}=\text{Age}_{\Lambda\text{CDM}}
	\label{eq:ageLCDM}
	\end{equation}
	Thus, we arrive at the familiar standard formula for calculating the age of the universe. By setting $a$ to 1 and inputting the estimates of $\Omega_{\text{m0}}$ and $H_0$, we can determine the current age of the universe, which is sensitive to the tension in $H_0$. 
\end{widetext}

\begin{figure}[h]
	\includegraphics[width=0.49\textwidth]{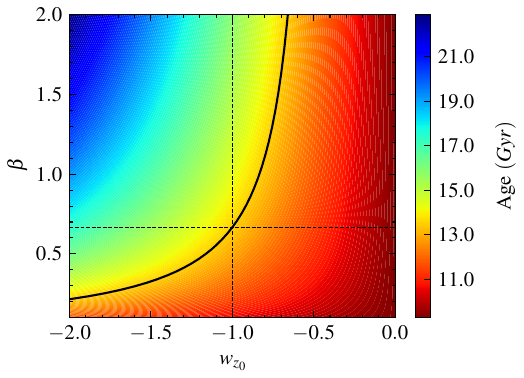}
	\caption{(colour online) Age of the universe in the GOHDE model with $H_0=70$ km/s/Mpc and $\Omega_{\text{m0}}=0.3$ for various combinations of $\beta$ and $w_{z_0}$. The solid line represents the $\Lambda$CDM estimate for $H_0=70$ km/s/Mpc and $\Omega_{\text{m0}}=0.3$ using Eq. (\ref{eq:ageLCDM}), which is $13.48~Gyr$.}
	\label{fig:age}
\end{figure}

We can now explore the influence of $\beta$ and $w_{z_0}$ on the age of the universe based on Eq. (\ref{eq:ageGOHDE}). For illustration purpose, we assume $H_0=70$ km/s/Mpc and $\Omega_{\text{m0}}=0.3$. In Fig. (\ref{fig:age}), we can see how different combinations of $\beta$ and $w_{z_0}$ affect the estimated age of the universe. Notably, when $w_{z_0}$ remains constant, a lower value of $\beta$ results in a significantly younger universe, while a higher $\beta$ leads to an older estimate compared to the $\Lambda$CDM model. The solid curve represents the universe's age in the $\Lambda$CDM model under the specified cosmological parameters, which is 13.48 billion years according to Eq. (\ref{eq:ageLCDM}). The actual age will depend on the estimate of  $H_0=$ and $\Omega_{\text{m0}}$, which according to the Planck 2018 release is $\sim13.8$ billion years \cite{aghanim2020planck}. 

Large-scale cosmological surveys can play a crucial role in constraining the values of both $w_{z_0}$ and the age of the universe. By doing so, we can establish limits on the parameter $\beta$.\\

\begin{tcolorbox}[colback=red!5!white,colframe=red!75!black]
	\textit{ Existing literature suggests that $\beta$ should be very close to $2/3$. This inference raises the possibility that the $\Lambda$CDM model might be the only viable and sensible solution within the framework of GOHDE with dark energy and matter.}
\end{tcolorbox}

To see this, let us move on to the actual parameter estimation promised earlier and perform statistical tests. If the data prefer a value close to $2/3$ for $\beta$ and $-1$ for $w_{z_0}$, then the GOHDE is practically indistinguishable from $\Lambda$CDM. 

\section{Observational Constrains \label{sec:5}}

Up to this point, we have analysed the GOHDE model by considering specific cosmological parameters within its parameter space. We have also made it more or less clear about the status of the $\Lambda$CDM model in the framework of the GOHDE model. This section uses various datasets and statistical techniques to focus on the GOHDE model's behaviour.

We perform MCMC (Markov Chain Monte Carlo) analyses for four different models to ensure a comprehensive analysis of each model and to estimate the model parameters \cite{Foreman-Mackey_2013}. This approach is motivated by the desire to minimise potential biases arising from unknown systematic or calibration errors that might be present in various analysis schemes. Some analyses, especially those specific to particular experiments, may involve unique software and calibration procedures not widely disclosed or understood by the public. Consequently, to maintain the fidelity of our analysis, we conduct MCMC analyses for the $\Lambda$CDM and $w$CDM models alongside the GOHDE models. This approach enables us to make faithful comparisons among the models within the scope of our analysis while accounting for any data-specific uncertainties that may be inherent to different datasets.

For simplicity and to focus on the critical aspects of our analysis, we will not account for radiation density, and we will exclusively examine the spatially flat cosmological scenario. While incorporating radiation density and spatial curvature could make the theoretical analysis more involved, the impact of these factors on our results is expected to be negligible due to the uncertainties in the data. The cornerstone of our analysis is the expression for the Hubble parameter, which serves as the foundation for the cosmological frameworks and defines all associated parameters. With the Hubble parameter as our starting point, we can make predictions about various observations and subject them to statistical tests.

In this analysis, we will explore four different cosmological models. The first two are the well-established $\Lambda$CDM and $w$CDM models. The equation we adopt for the $\Lambda$CDM model is
\begin{equation}
H=H_0\sqrt{\Omega_{\text{m0}}(1+z)^3+(1-\Omega_{\text{m0}})},
\end{equation}  
with two free parameter $H_0$ and $\Omega_{\text{m0}}$.
Similarly for the $w$CDM case we assume the value of $w$ to be constant and the respective expression takes the form,
\begin{equation}
H=H_0\sqrt{\Omega_{\text{m0}}(1+z)^3+(1-\Omega_{\text{m0}})(1+z)^{3(1+w)}},
\end{equation}
where we have three free parameter $H_0$, $\Omega_{\text{m0}}$ and $w$.

Further, we consider two versions of the GOHDE model. Initially, we didn't make any distinction between these two models, but we will now differentiate them and assign distinct names. The GOHDE model we've discussed has two additional parameters compared to the $\Lambda$CDM model: $\beta$ and $w_{z_0}$. For our first version, we will set $w_{z_0}=-1$ and treat $\beta$ as the extra free parameter. This version will be referred to as the ``GOHDE'' model. Next, we will consider a version in which $\beta$ and $w_{z_0}$ are free parameters. This model will be referred to as the ``$w$GOHDE'' model. Earlier in the text, we referred to both as GOHDE and explicitly mentioned the values of $\beta$ and $w_{z_0}$.
\begin{widetext}
	\noindent For the $w$GOHDE model we have
	\begin{equation}
	H=H_0\left\lbrace\frac{(z+1)^3 \left[2 \Omega_{\text{m0}}+3 \beta  w_{z_0} (\Omega_{\text{m0}} -1)
		\left(\frac{1}{z+1}\right)^{\frac{2 \Omega_{\text{m0}} }{\beta }+3
			w_{z_0} (\Omega_{\text{m0}} -1)} \right]}{3 \beta  w_{z_0}
		(\Omega_{\text{m0}} -1)+2 \Omega_{\text{m0}} }\right\rbrace^{1/2},
	\end{equation}
	with free parameters $H_0$, $\Omega_{\text{m0}}$, $\beta$ and $w_{z_0}$. Further, for GOHDE with fixed $w_{z_0}=-1$ we have,
	\begin{equation}
	H=H_0\left\lbrace\frac{(z+1)^3 \left[2 \Omega_{\text{m0}}+3 \beta (1-\Omega_{\text{m0}})
		\left(\frac{1}{z+1}\right)^{\frac{2 \Omega_{\text{m0}} }{\beta }+3
			(1-\Omega_{\text{m0}})} \right]}{3 \beta  
		(1-\Omega_{\text{m0}})+2 \Omega_{\text{m0}} }\right\rbrace^{1/2}.
	\end{equation}
	Where $H_0$, $\Omega_{\text{m0}}$ and $\beta$ are the free parameters. 
\end{widetext}
Defining the prior range is a critical step in any data analysis using the MCMC method. In this analysis, we have opted for a uniform prior values, which are as follows,
\begin{table}[h]
	\renewcommand{\arraystretch}{1.5}
	\begin{ruledtabular}
		\begin{tabular}{ccc}
			Parameter	& Model(s) &Prior range \\
			\hline
			$H_0$& All & [50,100]\\
			$\Omega_{\text{m0}}$& All  & [0.01,1]\\
			$M$	& All & [-25,-15]\\
			$w$ & $w$CDM  & [-2,0]\\
			$w_{z_0}$ 	& $w$GOHDE & [-2,0]\\
			$\beta$	& $w$GOHDE \& GOHDE & [0.1,2]\\
		\end{tabular}
	\end{ruledtabular}
	\caption{Prior values of all free parameters taken for the MCMC analysis.}
\end{table}
The prior values are selected with specific considerations. For $\beta$, we choose a range greater than zero to prevent potential singular behaviour at zero. The upper limit is three times that of the $\Lambda$CDM case. The priors for $H_0$ and $M$ are set in the $[50,100]$ and $[-25,-15]$, respectively, to avoid bias towards the values reported in the Hubble tension. The priors for $w$ and $w_{z_0}$ are identical and encompass extreme values compared to those found in the literature. Finally, the prior for the present matter density is chosen to exclude zero numerically. Now that we have the Hubble parameter expressions for each model and the prior, we can proceed to our analysis after discussing the data sets used. 

\subsection{Pantheon Type Ia Supernovae (SNe Ia) data} 

Type Ia supernovae (SNe Ia) are considered one of the most useful cosmic entities to explore the cosmic distance. Popularly known as the standard candles in cosmology, they provide one of the most reliable sources of distance measurement based on their luminosity. Several SNe Ia data have been compiled, starting with the milestone supernova project \cite{Riess_1998, Perlmutter1998}. A detailed list of each survey and the corresponding references are listed in another milestone compilation called the Pantheon sample \cite{Scolnic_2018}. The Pantheon sample is one of the most extensive data sets consisting of 1048 SNe Ia between redshift $ 0.01 < z < 2.3$. The successor of Pantheon is also available under the title Pantheon+ sample, which we have not used in this analysis \cite{Brout_2022}. Pantheon+ consists of 1550 SNe Ia, calibrated with the SH0ES estimates, which does put tension in the value of $H_0$ \cite{Riess_2022}. Since our purpose is not to address the tension first-hand, we shall keep the Pantheon+ sample aside and use the Pantheon sample alone.

Supernova data, in general, cannot put constraints on the value of $H_0$, as there is a degeneracy between $H_0$ and supernova absolute magnitude ($M$) in the expression used to fit the SNe light curves. It must be combined with other observations to estimate the value of $H_0$. Utilizing only Pantheon for reporting $H_0$ is impossible \cite{Brout_2022, Scolnic_2018}, and in its predecessor, the 580 Union2.1 dataset, $H_0$ is presumed \cite{suzuki2012hubble}. This data set has three main components: the apparent magnitude $(\mu)$, its respective redshift $(z)$ and the standard deviation in $\mu$. Thus, we aim to calculate the value of $\mu$ using the cosmological model of interest for the analysis. The apparent magnitude is given as,
\begin{equation}
\mu(z)=5\log_{10}\left[\frac{d_L(z)}{\text{Mpc}}\right]+M+25.
\label{eq:apparentmagnitude}
\end{equation}
Where $M$ is absolute magnitude, which needs to be calibrated using other data sets, $d_L$ is the luminosity distance given by the expression,
\begin{equation}
d_L(z)=c(1+z)\int_{0}^{z}\frac{1}{H(z')}dz'.
\label{eq:D_L}
\end{equation}
Using the Hubble parameter $H(z)$ at a specific redshift $z$, it is possible to estimate the luminosity distance $d_L$ and subsequently calculate the apparent magnitude $(\mu)$. This prediction is a crucial tool for assessing the model's significance compared to observational data.

\subsection{Observational Hubble Data}

The Observational Hubble Data (OHD) is a comprehensive collection of Hubble parameter measurements obtained from various sources, making it a valuable resource for cosmological analysis. Unlike a single dataset like Pantheon, OHD comprises a combination of correlated and independent estimations of the Hubble parameter at different redshifts. Specifically, it includes 31 data points derived from distance ladder estimations and 26 non-correlated data points based on baryonic acoustic oscillations (BAO). OHD consists of 57 data points, enabling a more comprehensive exploration of cosmological parameters. The complete list of data sources can be found in the reference \cite{SUDHARANI2023250}, which serves as our primary source for this dataset.

The redshift range covered by this dataset extends from $0.07$ to $2.36$, providing a broad scope for estimating the free parameters. This range also justifies dropping the radiation density parameter as little influence exists. When combined with the Pantheon dataset, OHD offers a powerful tool for accurately determining the value of the $H_0$, often achieving precision within 1 km/s/Mpc, as illustrated in Table 11 of \cite{Scolnic_2018}. The dataset also serves as a valuable means to explore the parameter space since it directly provides us with measurements of the Hubble parameter. (See subsection (\ref{subsec:quick}))

\subsection{CMB Shift parameter}
Utilizing the complete Cosmic Microwave Background (CMB) observations for comprehending the cosmological model is the ultimate test for any model. Nevertheless, conducting a thorough analysis from scratch is a formidable undertaking. One could reconfigure the Planck results series based on the new model, which would be interesting. However, this endeavour lies beyond the scope of the present manuscript.

Recognizing that CMB measurements are inherently dependent on a background cosmology is crucial. Almost all cosmological values are derived from quantities rooted in these background relations. While this may initially render these values unsuitable for testing other cosmological models, it is far from a dead-end proposition. It becomes feasible to employ these derived parameters, along with their associated error bars, once one can establish a reasonably accurate expectation for the distribution arising from the assumption of the primordial power spectrum. This specific aspect has been extensively explored in  \cite{refId0Elgaroy}. They demonstrated that relying solely on the CMB shift parameter may not be adequate. In their analysis, they incorporated both the shift parameter and the position of the first acoustic peak in the multipole space to enhance the constraints on the dark energy model. Thus, integrating the shift parameter with other observational data can alleviate the associated challenges, ultimately providing a more meaningful sense of the constraints.

Here, the CMB shift parameter is given as,

\begin{equation}
\mathcal{R}=\sqrt{\Omega_{\text{m0}}}\int_{0}^{z_r}\frac{1}{h(z)}dz.
\end{equation} 

Where $z_r$ represents the recombination redshift, this measurement can impose rigorous constraints on the value of $\Omega_{\text{m0}}$. For our purposes, we adopt the value derived from the Planck 18 analysis as documented in \cite{Chen_2019}.

\begin{equation*}
\mathcal{R}=1.7502\pm 0.0046 \text{ with } z_r=1089.92
\end{equation*}

It is essential to note that while the relationship between the shift parameter and the acoustic peak is linear, they are not degenerate and can complement each other. Another significant quantity, the drag epoch ($r_s$), also holds this relation, elaborated in \cite{10.1093/mnras/stz1966}. With this insight, we have chosen to employ the CMB shift parameter in conjunction with both Baryon Acoustic Oscillation (BAO) and Quasi-Stellar Object (QSO) datasets, in contrast to the combination recommended in \cite{refId0Elgaroy}. Furthermore, we have incorporated the OHD and Pantheon datasets into separate data combinations for a comprehensive analysis.

\subsection{BAO data}

Baryon Acoustic Oscillations (BAO) data has emerged as a staple observational tool for constraining cosmological models. These data primarily originate from surveys of the large-scale structure power spectrum, such as the SDSS-III with DR12 galaxy sample \cite{10.1093/mnras/stx721}, and are comprehensively documented in \cite{10.1093/mnras/stab1373}. In our analysis, we focus on data points corresponding to two specific parameters: the transverse comoving distance $D_M(z)$, which coincides with $D_c(z)$ in a flat universe, and the volume-averaged angular diameter distance $D_v(z)$. We temporarily omit the angular diameter distance $D_A(z)$, as it will be incorporated when integrating the QSO data. Additionally, we exclude the Hubble values since they are already encompassed within the OHD dataset. It's worth noting that all of these distance parameters are scaled by the values of $\mathcal{R}$ and $r_s$, enabling us to integrate the CMB shift parameter mentioned earlier into our analysis. This integration enhances our ability to achieve a more refined constraint on the cosmological model. The relevant expressions are, 
\begin{align}
D_M(z)=&D_c(z)=c\int_{0}^{z}\frac{1}{H(z')}dz',\\
D_V(z)=&\left[\frac{cz}{H(z)}D_M^2(z)\right]^{1/3}.
\end{align}
Here, we use data points from the sources cited in \cite{10.1093/mnras/stab1373, 10.1093/mnras/stz1966, 10.1093/mnras/stx721}.

\subsection{QSO data}
We also incorporate another dataset derived from ultra-compact structures in radio sources. This dataset consists of 120 data points corresponding to angular sizes and redshifts observed in intermediate-luminosity quasars spanning a redshift range from $0.46$ to $2.76$, as detailed in \cite{refId0Ultra}. These quasars exhibit minimal dependence on redshift and intrinsic luminosity when observed at 2.29 GHz, effectively establishing a standardised ruler with a linear size of approximately $11.03\pm0.25$ pc.

Utilising this ruler, we can assess the validity of our cosmological model. The relationship connecting the angular size $(\theta)$, linear length scale $(l_m)$, and angular diameter distance $(D_A)$ for a given redshift is defined as follows:

\begin{equation}
\theta(z)=\frac{l_m}{D_A(z)}.
\end{equation} 

Here, we utilise the angular diameter distance ($D_A(z)$), which is defined as the luminosity distance divided by $(1+z)^2$. The luminosity distance is determined using Equation (\ref{eq:D_L}), similar to its application in the case of SNe Ia.

In line with the methodology and dataset outlined in \cite{refId0Ultra}, we introduce an additional 10\% error to the angular size standard deviation to account for any potential other uncertainties.

Considering that both the QSO and BAO datasets inherently incorporate the drag redshift, we anticipate they synergise effectively with the CMB Shift parameter, enhancing our ability to estimate our free parameters. Consequently, we opt for the QSO and BAO combinations instead of the acoustic scale parameter mentioned in \cite{refId0Elgaroy}. These strategies are versatile and applicable to the comprehensive study of various cosmological models.

\subsection{Data combinations}

Each of the datasets mentioned above cannot independently constrain the model parameters. For instance, it is impossible to determine the value of $H_0$ using either the Pantheon dataset or the CMB Shift parameter alone. Therefore, in our analysis, we consider four distinct data combinations, collectively referred to as D1, D2, D3, D4, and the Full Data set. These combinations are structured as follows:
\begin{table}
	\renewcommand{\arraystretch}{1.5}
	\begin{ruledtabular}
		\begin{tabular}{cc}
			Data Set Name & Data Combination used \\
			\hline
			Full & OHD + Pantheon + CMB + BAO + QSO \\
			D1 & OHD + Pantheon + BAO \\
			D2 & Pantheon + QSO \\
			D3 & OHD + Pantheon + BAO + CMB\\
			D4 &  Pantheon + QSO + CMB\\
		\end{tabular} 
	\end{ruledtabular}
	\label{tab:DataSets}
	\caption{Cosmological data set combination used for the analysis. }
\end{table}
Within these combinations, the Full Data set incorporates the entirety of the previously introduced data. On the other hand, data sets D1 and D2 exclude the CMB shift parameter. To complement these analyses, we replicate these combinations while including the CMB data in D3 and D4, respectively. Regardless of the specific data combination, the Pantheon dataset is consistently included in all analyses. The final estimate of the free parameter is taken from the Full set.

\subsubsection*{\textbf{Best fit values}}

With the essential tools at our disposal, we perform parameter estimation by minimising the $\chi^2$ function,
\begin{equation}
\chi^2=\sum_{i=1}^N\left(\frac{A^{\text{theory}}_i-A^{\text{observe}}_i}{\sigma_i}\right)^2,
\end{equation}
by employing the robust MCMC method. In the above expression, $N$ corresponds to the number of data points. The best fit values for each free parameter are meticulously outlined in TABLE (\ref{tab:BestFits}). Generally, the best fit parameters tend to favour values close to those estimated in the $\Lambda$CDM model.

Let us look at the behaviour exhibited by each model when subjected to different datasets. When utilising the Full dataset, the best-fit estimate for the GOHDE and $w$GOHDE models indicate a preference for a value of $\beta$ close to 2/3. This observation strongly suggests that $\beta$ should be approximately in the vicinity of $2/3$ based on current observations. Furthermore, when examining the present value of the equation of state parameter, denoted as $w_{z_0}$ in the context of the $w$GOHDE model, we find it closely approximating -1. This finding further supports our assertion that the $\Lambda$CDM model represents the most favourable particular case within the GOHDE framework. 

Notably, the model provides a Hubble parameter estimate that aligns closely with the results from the Planck mission. Additionally, the value of parameter $M$ is consistent with the constraints derived in a recent study documented in \cite{PhysRevD.107.063513}. However, it's essential to acknowledge that it still exhibits tension with the SH0ES estimate presented by Riess et al. in 2022 \cite{Riess_2022}. In fact, the tension observed in the Hubble constant ($H_0$) mirrors the tension in the parameter $M$, which is extensively discussed in \cite{PhysRevD.105.063524}. Thus, the Hubble tension is not automatically resolved unless we forcefully set bound to the free parameters.

\begin{table*}[t]
	\renewcommand{\arraystretch}{1.68}
	\begin{ruledtabular}
		\begin{tabular}{lllllll}
			Data	  &  	Model   &   $\Omega_m$   &   $M$   &   $H_0$   &   $w$ or $w_{z_0}$   &   $\beta$ \\ 
			\hline
			\multirow{4}{*}{\rotcell{Full}}	  &  	$\Lambda$CDM  &   $  0.2943^{+0.0063}_{-0.0065}    $   &   $-19.401^{+0.012}_{-0.013}$   &   $68.45^{+0.43}_{-0.43}$   &   --   &  --  \\ 
			&  	$w$CDM  &   $  0.2923^{+0.0070}_{-0.0063}    $   &   $-19.395^{+0.015}_{-0.014}$   &   $68.80^{+0.61}_{-0.57}$   &   $-1.023^{+0.024}_{-0.025}$   &  --\\ 
			&  	GOHDE  &   $  0.2875^{+0.0081}_{-0.00.0}    $   &   $-19.394^{+0.016}_{-0.014}$   &   $68.79^{+0.57}_{-0.51}$   &  --    &   $  0.6786^{+0.0095}_{-0.0097}    $ \\ 
			&  	$w$GOHDE  &   $  0.2858^{+0.0102}_{-0.0097}    $   &   $-19.393^{+0.014}_{-0.015}$   &   $68.77^{+0.60}_{-0.60}$   &   $-0.988^{+0.042}_{-0.044}$   &   $0.686^{+0.048}_{-0.044}$ \\ 
			\hline
			\multirow{4}{*}{\rotcell{D1}}	  &  	$\Lambda$CDM   &   $0.2846^{+0.0099}_{-0.0105} $   &   $-19.395^{+0.015}_{-0.015}$   &   $68.75^{+ 0.57}_{- 0.57}$   &   --  &  --\\ 
			&  	$w$CDM   &   $0.281^{+0.014}_{-0.015}$   &   $-19.395^{+0.014}_{-0.016}$   &   $68.69^{+0.59}_{-0.63}$   &   $-0.981^{+0.047}_{-0.050}$   &   -- \\ 
			&  	GOHDE   &   $0.299^{+0.025}_{-0.027}$   &   $-19.396^{+0.015}_{-0.015}$   &   $68.59^{+0.64}_{-0.56}$   &  --   &   $0.71^{+0.14}_{-0.11}$\\ 
			&  	$w$GOHDE   &   $0.573^{+0.082}_{-0.184}$   &   $-19.397^{+0.016}_{-0.014}$   &   $68.64^{+0.60}_{-0.59}$   &   $-1.44^{+0.39}_{-0.37}$   &   $1.33^{+0.28}_{-0.44}$
			\\ 
			\hline
			\multirow{4}{*}{\rotcell{D2}}	  &  	$\Lambda$CDM   &   $0.285^{+0.013}_{-0.013}$   &   $-19.355^{+0.043}_{-0.038}$   &   $70.0^{+1.5}_{-1.3}$   &   --   &   -- \\ 
			&  	$w$CDM   &   $0.354^{+0.031}_{-0.035}$   &   $-19.382^{+0.045}_{-0.042}$   &   $69.7^{+1.4}_{-1.4}$   &   $-1.23^{+0.14}_{-0.14}$   &   -- \\ 
			&  	GOHDE   &   $0.226^{+0.052}_{-0.050}$   &   $-19.375^{+0.046}_{-0.042}$   &   $69.7^{+1.5}_{-1.3}$   &   --   &   $0.35^{+0.20}_{-0.15}$ \\ 
			&  	$w$GOHDE   &   $0.531^{+0.085}_{-0.179}$   &   $-19.374^{+0.043}_{-0.044}$   &   $69.7^{+1.5}_{-1.3}$   &   $-1.39^{+0.24}_{-0.47}$   &   $0.74^{+0.31}_{-0.32}$ \\ 
			\hline
			\multirow{4}{*}{\rotcell{D3}}	  &  	$\Lambda$CDM   &   $ 0.2939^{+0.0065}_{-0.0064}  $   &   $-19.405^{+0.014}_{-0.012}$   &   $68.35^{+0.44}_{-0.42}$   &   --   &   -- \\ 
			&  	$w$CDM   &   $ 0.2922^{+0.0069}_{-0.0062}  $   &   $-19.398^{+0.015}_{-0.015}$   &   $68.72^{+0.60}_{-0.59}$   &   $-1.022^{+0.024}_{-0.025}$   &   -- \\ 
			&  	GOHDE   &   $ 0.2866^{+0.0086}_{-0.0084}  $   &   $-19.396^{+0.015}_{-0.015}$   &   $68.70^{+0.59}_{-0.50}$   &   --   &   $ 0.679^{+0.0096}_{-0.0096} $ \\ 
			&  	$w$GOHDE   &   $ 0.2859^{+0.0094}_{-0.0106}  $   &   $-19.398^{+0.016}_{-0.014}$   &   $68.62^{+0.61}_{-0.57}$   &   $-0.986^{+0.042}_{-0.043}$   &   $0.689^{+0.048}_{-0.045}$ \\ 
			\hline
			\multirow{4}{*}{\rotcell{D4}}	  &  	$\Lambda$CDM   &   $ 0.2959^{+0.0074}_{-0.0068}$   &   $-19.363^{+0.038}_{-0.039}$   &   $69.6^{+1.2}_{-1.3}$   &   --   &   -- \\ 
			&  	$w$CDM   &   $0.2929^{+0.0076}_{-0.0070} $   &   $-19.352^{+0.039}_{-0.040}$   &   $70.2^{+1.4}_{-1.3}$   &   $-1.035^{+0.026}_{-0.025}$   &   -- \\ 
			&  	GOHDE   &   $0.287^{+0.012}_{-0.011}$   &   $-19.353^{+0.042}_{-0.040}$   &   $70.1^{+1.4}_{-1.4}$   &   --   &   $0.678^{+0.012}_{-0.012} $ \\ 
			&  	$w$GOHDE   &   $0.329^{+0.022}_{-0.024}$   &   $-19.377^{+0.042}_{-0.045}$   &   $69.7^{+1.4}_{-1.4}$   &   $-1.17^{+0.11}_{-0.12}$   &   $0.507^{+0.086}_{-0.067}$ \\ 
		\end{tabular}
	\end{ruledtabular}
	\caption{Best fit values for four distinct models ($\Lambda$CDM, $w$CDM, GOHDE, and $w$GOHDE) across five data combinations (Full, D1, D2, D3, and D4). We have consolidated both $w$ and $w_{z_0}$ within the same column, even though the parameter for the $w$CDM model is denoted as $w$ and that for the $w$GOHDE model is referred to as $w_{z_0}$.}
	\label{tab:BestFits}
\end{table*}
Let us now examine the best fit estimates based on various data combinations. Data sets D1 and D2 do not incorporate the CMB shift parameter. We observe that the constraints on the values of $w_{z_0}$ and $\beta$ experience notable enhancement when the CMB shift parameter is introduced in sets D3 and D4. For example, within the D2 data set, comprising Pantheon and QSO data, the estimation for $w_{z_0}$ stands at $-1.39^{+0.24}_{-0.47}$. Upon inclusion of the CMB Shift parameter in the D4 set, this constraint refines to $-1.17^{+0.11}_{-0.12}$. Similarly, in the case of data sets D1 and D3, the constraints on $w_{z_0}$ exhibit a substantial improvement, transitioning from $-1.44^{+0.39}_{-0.37}$ to a much tighter constraint of $-0.986^{+0.042}_{-0.043}$.

\begin{figure}[h]
	\includegraphics[width=0.48\textwidth]{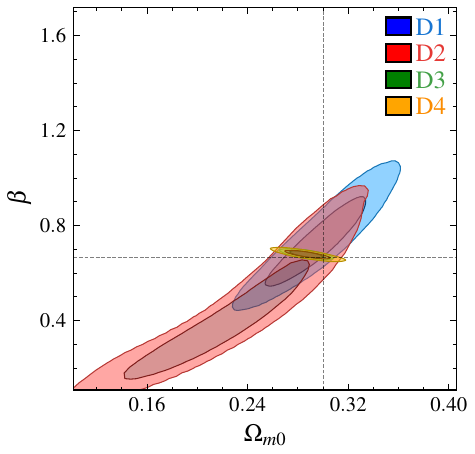}
	\caption{(colour online) Confidence plot between $\beta$ and $\Omega_{\text{m0}}$ for GOHDE under different data set combination. The Full data combination is excluded as the contours are very close to D3 set.}
	\label{fig:Om_b1}
\end{figure}

\begin{figure}[h]
	\includegraphics[width=0.48\textwidth]{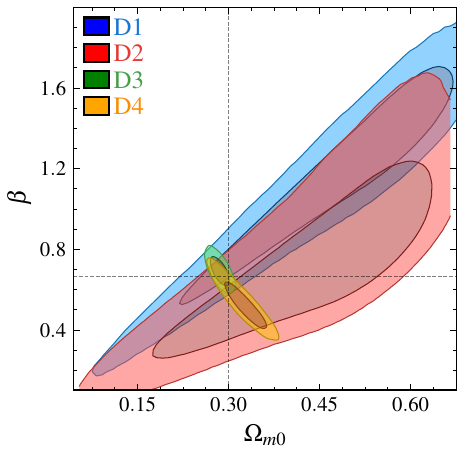}
	\caption{(colour online) Confidence plot between $\beta$ and $\Omega_{\text{m0}}$ for $w$GOHDE under different data set combination. The Full data combination is excluded as the contours are very close to D3 set.}
	\label{fig:Om_b2}
\end{figure}

\begin{figure}[h]
	\includegraphics[width=0.48\textwidth]{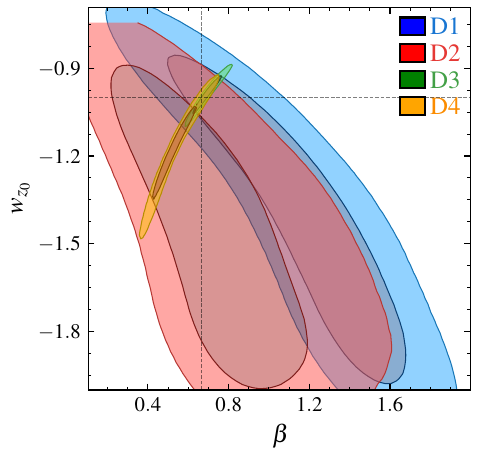}
	\caption{(colour online) Confidence plot of $w_{z_0}$ with $\beta$  for the $w$GOHDE model under different data set combination.The Full data combination is excluded as the contours are very close to D3 set.}
	\label{fig:wGOHDEbwO1}
\end{figure}

\begin{figure}[h]
	\includegraphics[width=0.48\textwidth]{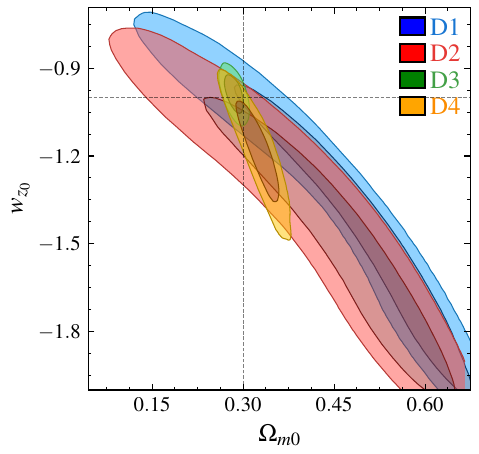}
	\caption{(colour online) Confidence plot of $w_{z_0}$ with  $\Omega_{\text{m0}}$ for the $w$GOHDE model under different data set combination. The Full data combination is excluded as the contours are very close to D3 set.}
	\label{fig:wGOHDEbwO2}
\end{figure}

The underlying reason behind this improved outcome upon adding CMB shift parameter data pertains to the matter density. It becomes evident that in the absence of the shift parameter, the estimated matter density tends to be higher, typically around $\gtrsim0.5$. Such a high matter density is generally less favoured within the context of cosmological models. The crucial point here is that CMB observations impose stringent constraints on the value of $\Omega_{\text{m0}}$, which in turn leads to the enhancement of constraints on $w_{z_0}$ and $\beta$ within the $w$GOHDE model. Additionally, several general characteristics of QSO data become evident when examining the estimates, such as a notably high standard deviation in $H_0$, consistent with results tabulated in table (I) in \cite{Lenart_2023}. We can achieve considerably more stringent constraints by considering data combinations than when we rely solely on individual data sets. The enhancement in constraints becomes apparent when incorporating the CMB shift parameter, as evident in the confidence plots in Figs. (\ref{fig:Om_b1}) to (\ref{fig:wGOHDEbwO2}). In all these plots, the dotted lines corresponding to the $\Lambda$CDM model align statistically close with the best-fit estimates, particularly in the presence of the CMB shift parameter. This highlights the deep correlation between the free parameters of $w$GOHDE or GOHDE with the matter density.

The feasibility of modelling dark energy through the holographic principle hinges on the intricate behaviour of the matter density. The evolution of HDE, both in general contexts and specifically within our analysis, exhibits a strong dependence on matter dynamics. It's worth noting that this scenario would have presented a radically different landscape if we were dealing with radiation instead of matter.

These findings suggest that the holographic principle presents itself as a natural way to set constraints on total energy density, extending beyond its traditional application solely to dark energy density. It is important to note that while modelling dark energy based on the holographic principle holds promise, it may not yield results as straightforward or fruitful as initially anticipated. We shall move the statistical analysis to see whether our conclusions remain true.

\section{Statistical analysis}
To gain a more profound understanding of the behaviour of data combinations and the nature of the models, it is advantageous to estimate specific statistical metrics beyond merely examining the best-fit values. Several methodologies exist for comparing models; we outline a few of these methods in this section. We will subsequently assess the goodness of fit of our models of interest based on these statistical estimates. Let's begin by examining these quantifiers one by one.

\subsubsection*{\textbf{$\chi^2$ distribution and degrees of freedom}}
The $\chi^2$ quantifies the deviation of the model from the observations while considering the respective errors associated with each data point in a dataset comprising $N$ observations, each with individual standard deviations $\sigma_i$. Mathematically, $\chi^2$ is defined as,

\begin{equation}
\chi^2 = \sum_{i=1}^N \left(\frac{A^{\text{theory}}_i - A^{\text{observe}}_i}{\sigma_i}\right)^2.
\end{equation}

Here, $A$ represents the quantity of interest. The best-fit estimates discussed earlier correspond to the parameter values that minimize this $\chi^2$. In an ideal scenario where a model perfectly explains all data points, the $\chi^2$ value would be zero. Consequently, a smaller $\chi^2$ indicates a better fit of the model to the data. As our parameter estimation relies on the $\chi^2$ minimization method, we can easily extract the $\chi^2$ distribution from the converged MCMC chain. However, it's important to note that $\chi^2$ alone does not account for the penalty of introducing additional parameters into a model to achieve a better fit. As an initial step towards comparing models, we can use the $\chi^2$ degrees of freedom, defined as,

\begin{equation}
\chi^2_{\text{\tiny{dof}}} = \frac{\chi^2}{N - N_p}.
\end{equation}

Here, $N_p$ represents the number of free parameters in the model. In the case of a realistic model, the value of $\chi^2_{\text{dof}}$ tends to approach unity. However, while $\chi^2_{\text{dof}}$ penalizes adding extra parameters in the model, this approach is only an initial step.

\subsubsection*{ \textbf{Akaike Information Criterion (AIC)\\ and \\Bayesian Information Criterion (BIC)}}
To address the challenges of model selection, H. Akaike introduced an information criterion, later known as the Akaike Information Criterion (AIC), in 1974 \cite{1100705}, while G. Schwarz proposed the Bayesian Information Criterion (BIC) in 1978 \cite{10.1214/aos/1176344136}. These two approaches are valuable tools for picking a suitable model. While AIC primarily focuses on the number of free parameters, BIC is grounded in the principles of Bayesian estimator. While similarities exist between the two, BIC prefers a lower dimensional model, making it a more favourable choice for model selection over AIC. The expressions for AIC and BIC are,

\begin{align}
\text{AIC} &= N\log\left(\frac{\chi^2}{N}\right) + 2N_p,\\
\text{BIC} &= N\log\left(\frac{\chi^2}{N}\right) + \left[\frac{1}{2}\log(N)\right]2N_p.
\end{align}

Qualitatively, both AIC and BIC appear pretty similar, with the notable distinction of having a multiplicative factor $\frac{1}{2}\log(N)$ applied to the number of free parameters in BIC \cite{10.1214/aos/1176344136}, which gives BIC its advantage over the other.

\subsubsection*{\textbf{Bayesian Analysis}}
Up to this point, the statistical quantifiers we've discussed have not explicitly considered the prior range of the analysis. Since introducing an explicit prior can substantially impact both the best-fit values and the significance of the model, we conduct a Bayesian analysis between the models. This relies on the Bayes theorem, which says,
\begin{equation}
\text{Posterior probability } (\mathcal{P})=\frac{\text{Likelihood } (\mathcal{L}) \times \text{Prior } (\pi)}{\text{Evidence of data } (\mathcal{E})} .
\end{equation}
Now, the Bayesian evidence $\mathcal{E}$ for the data $D$ and model $\mathcal{M}$ in a given parameter space $p$ takes the from,
\begin{equation}
\mathcal{E}(D|\mathcal{M})=\int dp \mathcal{L}(D|p,\mathcal{M}) \pi(p,\mathcal{M}).
\end{equation}
Now the Bayes factor between model $\mathcal{M}_i$ and $\mathcal{M}_j$ is defined as,
\begin{equation}
B_{ij}=\frac{\mathcal{E}(D|\mathcal{M}_i)}{\mathcal{E}(D|\mathcal{M}_j)}.
\end{equation}
The above quantity gives a numerical estimate of how much evidence a model shows over another. Here, we follow the Jeffreys scale named after Harold Jeffreys \cite{jeffreys1998theory} given in the following table to compare the models. 
\begin{table}[h]
	\renewcommand{\arraystretch}{1.5}
	\begin{ruledtabular}
		\begin{tabular}{ccc}
			$B_{ij}$   &   $\ln{B_{ij}}$  &  Evidence  \\
			\hline
			$0\leq B_{ij}<1$    &  $B_{ij}<0$  &  Negative       \\
			$1\leq B_{ij}<3$    &  $0\leq B_{ij}<1.1$  &  Weak       \\
			$3\leq B_{ij}<20$   &  $1.1\leq B_{ij}<3$    &  Definite   \\
			$20\leq B_{ij}<150$  &  $3\leq B_{ij}<5$    &  Strong     \\
			$150\leq B_{ij}$  &  $5\leq B_{ij}$    &  Very Strong\\
		\end{tabular}
	\end{ruledtabular}
	\caption{The Jeffreys' scale of Bayesian evidence  \cite{jeffreys1998theory}.}
	\label{tab:Jeffreys}
\end{table}
Unless $B_{ij}$ goes beyond $3$, the model $i$ is not taken seriously.
\begin{table*}[t]
	\renewcommand{\arraystretch}{1.5}
	\begin{ruledtabular}
		\begin{tabular}{llllllll}
			Data	 & 	Model  &  $\chi^2$  &  $\chi^2_{\text{\tiny dof}}$  &  AIC  &  BIC  &  $\Delta$AIC & $\Delta$BIC\\
			\hline
			\multirow{4}{*}{\rotcell{Full}}	 & 	$\Lambda$CDM  &  1426.51  &  1.1616  &  187.45  &  202.79  &  0 & 0 \\
			& 	$w$CDM  &  1425.59  &  1.1618  &  188.67  &  209.13  &  1.21 & 6.34 \\
			& 	GOHDE  &  1424.81  &  1.1612  &  187.99  &  208.45  &  0.54 & 5.66 \\
			& 	$w$GOHDE  &  1424.92  &  1.1622  &  190.09  &  215.67  &  2.63 & 12.88 \\
			\hline
			\multirow{4}{*}{\rotcell{D1}}	 & 	$\Lambda$CDM  &  1071.41  &  0.9678  &  -33.28  &  -18.24  &  0&  0 \\
			& 	$w$CDM  &  1071.29  &  0.9686  &  -31.40  &  -11.35  &  1.87 & 6.89 \\
			& 	GOHDE  &  1071.26  &  0.9685  &  -31.435  &  -11.38  &  1.85 & 6.86 \\
			& 	$w$GOHDE  &  1071.25  &  0.9695  &  -29.448  &  -4.38  &  3.84 & 13.86 \\
			\hline
			\multirow{4}{*}{\rotcell{D2}}	 & 	$\Lambda$CDM  &  1387.81  &  1.1912  &  207.402  &  222.591  &  0 & 0 \\
			& 	$w$CDM  &  1384.53  &  1.1894  &  206.639  &  226.892  &  -0.762 & 4.301 \\
			& 	GOHDE  &  1384.54  &  1.1895  &  206.645  &  226.897  &  -0.757 & 4.306 \\
			& 	$w$GOHDE  &  1384.52  &  1.1905  &  208.630  &  233.945  &  1.228&  11.354 \\
			\hline
			\multirow{4}{*}{\rotcell{D3}}	 & 	$\Lambda$CDM  &  1073.19  &  0.96859  &  -32.46  &  -17.4  &  0 & 0 \\
			& 	$w$CDM  &  1072.27  &  0.96862  &  -31.42  &  -11.4  &  1.04 & 6 \\
			& 	GOHDE  &  1071.38  &  0.96783  &  -32.34  &  -12.3  &  0.12&  5.1 \\
			& 	$w$GOHDE  &  1071.27  &  0.96860  &  -30.45  &  -5.4  &  2.01 & 12 \\
			\hline
			\multirow{4}{*}{\rotcell{D4}}	 & 	$\Lambda$CDM  &  1388.9  &  1.1912  &  207.5  &  222.73  &  0 & 0 \\
			& 	$w$CDM  &  1386.9  &  1.1905  &  207.8  &  228.11  &  0.32 & 5.38 \\
			& 	GOHDE  &  1387.9  &  1.1914  &  208.7  &  228.97  &  1.18 & 6.24 \\
			& 	$w$GOHDE  &  1384.6  &  1.1895  &  207.9  &  233.21  &  0.35 & 10.48 \\
		\end{tabular}
	\end{ruledtabular}
	\caption{Comparison of four different models ($\Lambda$CDM, $w$CDM, GOHDE and $w$GOHDE) based on five data combinations (Full, D1, D2, D3 and D4) using $\chi^2$, $\chi^2_{\tiny dof}$, AIC, BIC, $\Delta$AIC and $\Delta$BIC values. }
	\label{tab:ModelStat}
\end{table*}
\subsubsection*{\textbf{Statistical Inferences}}

\begin{figure}[b]
	\includegraphics[width=0.48\textwidth]{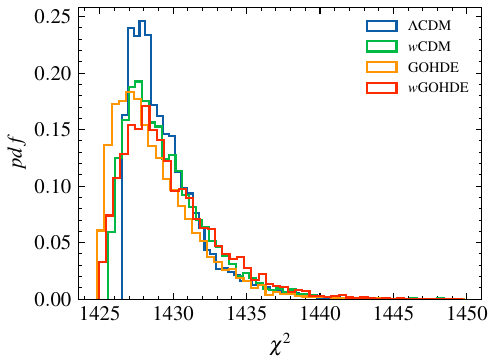}
	\caption{(colour online) $\chi^2$ distribution based on the converged MCMC chains for Full data combination.}
	\label{fig:chidist}
\end{figure}

Figure (\ref{fig:chidist}) displays histograms representing the $\chi^2$ distribution derived from the converged MCMC chain for all four models when considering Full data combinations. Notably, the $\chi^2$ distributions for the GOHDE and $w$GOHDE models exhibit substantial overlap with that of the $\Lambda$CDM model and are virtually indistinguishable from the $w$CDM model. This observation suggests that the data does not favour the GOHDE and $w$GOHDE models over the other concordance models. However, let us try to extract the details.

\begin{figure*}
	\subfloat[D1 data combination]{\includegraphics[width=0.48\textwidth]{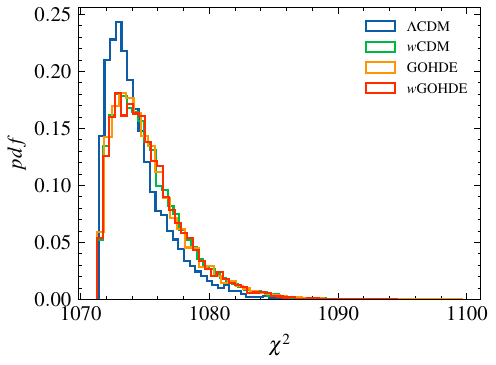}}
	\subfloat[D3 data combination]{\includegraphics[width=0.48\textwidth]{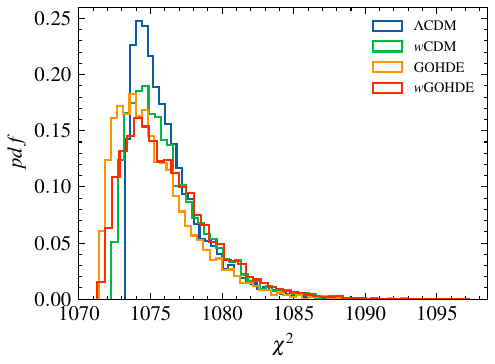}}\\
	\subfloat[D2 data combination]{\includegraphics[width=0.48\textwidth]{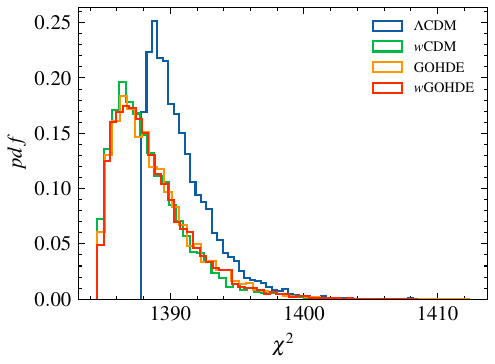}}\subfloat[D4 data combination]{\includegraphics[width=0.48\textwidth]{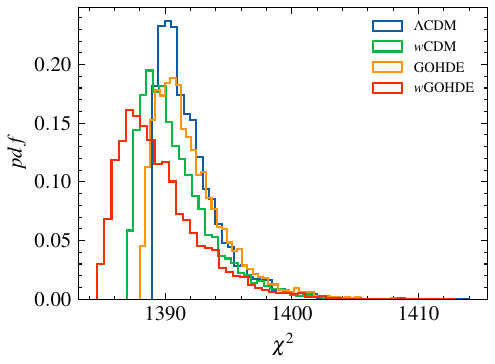}}
	\caption{(colour online) $\chi^2$ distribution based on the converged MCMC chains for various data combinations.}
	\label{fig:chidist2}
\end{figure*}
This does not mean that model $i$ is terrible; this indicates that the data does not show any preference towards model $i$ as opposed to model $j$. Thus, weak evidence suggests that data only partially distinguish between the models. However, negative evidence indicates that model $i$ appears less preferred over $j$. Although this method sounds like a robust classifier, it has some downfalls. A detailed critique on using this approach in cosmology is made in \cite{SavvasNesseris_2013}. Some notable reasons are the dependence on the prior value, distinct free parameters, etc. Hence, the Bayes factor is less decisive than AIC or BIC. Still, it can give reasonable information about the model that the data prefers.

\begin{table*}[t]
	\renewcommand{\arraystretch}{1.5}
	\begin{ruledtabular}
		\begin{tabular}{crrrrrr}
			Data	&  GOHDE / &  $w$GOHDE / &  $w$GOHDE / &  GOHDE / &  $w$GOHDE / &  $w$CDM /\\
			& $\Lambda$CDM &  $\Lambda$CDM &  GOHDE &$w$CDM &  $w$CDM &  $\Lambda$CDM \\
			\hline
			Full&\textbf{1.0004} &0.3463 & 0.3461 &1.5691 &0.5431 & 0.6376 \\
			D1& 0.4787&0.2192 & 0.4578 & 1.0330&0.4729 & 0.4634 \\
			
			D2 &2.0742&0.9721&0.4687 & 0.9914&0.4646&2.0922 \\
			
			D3	&1.0773& 0.4105& 0.3811 &1.6582& 0.6319&0.6497 \\
			
			D4	&0.6760  & 1.3589 &2.0102 &0.6114  & 1.2291 &1.1056 \\
		\end{tabular}
	\end{ruledtabular}
	\caption{Bayes factor $(B_{ij})$ between four different models ($\Lambda$CDM, $w$CDM, GOHDE and $w$GOHDE) based on five data combinations (Full, D1, D2, D3 and D4).}
	\label{tab:ModelBayes}
\end{table*}

TABLE (\ref{tab:ModelStat}) presents a comprehensive analysis of statistical quantifiers for various models across different data set combinations. Initially, it may seem that $\Lambda$CDM is not optimal for effectively explaining the data, particularly based on the $\chi^2$ minimization alone. However, this observation can be attributed to additional parameters that were not considered, and even when considering $\chi^2$ degrees of freedom, sufficient distinction between models is not achieved. We have computed various statistical parameters to address this limitation, including AIC, BIC, $\Delta$AIC, and $\Delta$BIC. A consistent trend across all these quantifiers is that the most favoured model is the one exhibiting the lowest values for $\chi^2$, $\chi^2_{\text{\tiny{dof}}}$, AIC, and BIC. Furthermore, $\Delta$AIC and $\Delta$BIC emphasize the effectiveness of BIC in penalizing models with additional free parameters. While we refrain from debating the credibility of AIC vs. BIC, we have estimated both for model comparison. In alignment with \cite{10.1214/aos/1176344136}, our findings indicate that BIC effectively distinguishes between models compared to AIC. Notably, BIC remains conservative even when AIC inclines toward models with extra free parameters.

After scrutinizing TABLE (\ref{tab:ModelStat}), several significant observations come to light. Firstly, $\chi^2$, $\chi^2_{\text{\tiny{dof}}}$, and AIC appear to be less effective for model comparison, a point of interest for statisticians. Conversely, when we turn our attention to BIC and $\Delta$BIC, it becomes apparent that $\Lambda$CDM consistently emerges as the preferred model, with others trailing behind. Although slight numerical variations may arise depending on the data combinations, the overarching trend remains consistent.

A second noteworthy finding is that $w$GOHDE does not appear to be the optimal choice, primarily due to its higher number of free parameters than the other three models. The BIC values further underscore this trend. However, it's worth noting that $w$GOHDE appears to perform better according to $\chi^2$. This observation highlights the importance of considering free parameters, as BIC does, during the model selection process.

Lastly, the similarity observed between the GOHDE and $w$CDM models is striking. Both models possess different yet identical numbers of free parameters, as evident in the $\Delta$BIC estimates. A closer examination of AIC suggests these models may perform comparably well compared to $\Lambda$CDM. However, this observation is corrected when using BIC as the information criterion. The similarity in BIC values between GOHDE and $w$CDM reveals a strong correlation between the free parameters $\beta$ and $w$. In the case of the $w$CDM model, $w$ represents the equation of state of dark energy throughout its evolution and is also the present value of the dark energy equation of state. Thus, this correlation extends to $\beta$ and $w_{z_0}$. This explains why $w$GOHDE performs less favourably compared to the other models, with BIC effectively penalizing it for overlooking such correlations. The confidence plots in Figs. (\ref{fig:wGOHDEbwO1}) and (\ref{fig:wGOHDEbwO2}) further support these observations. 

Up to this point, our analysis has yet to account for prior values, and it is clear that $\Lambda$CDM emerges as the best fit among all models. We calculate the Bayes factor for various model combinations to incorporate the influence of prior values and demonstrate the evidence favouring one model over another. Although the Bayes factor may not be the most suitable estimator, handling prior values properly makes it appropriate. 

In TABLE (\ref{tab:ModelBayes}), we present the Bayes factors for $\Lambda$CDM, $w$CDM, GOHDE, and $w$GOHDE based on all five data combinations (Full, D1, D2, D3, and D4). According to the Jeffreys scale in TABLE (\ref{tab:Jeffreys}), all models exhibit weak or negative evidence against the concordance $\Lambda$CDM. A particularly notable observation is the Bayes factor of GOHDE compared to $\Lambda$CDM with the Full dataset, which is almost unity. This indicates the data provides nearly identical evidence for $\Lambda$CDM and GOHDE. The Bayesian inference analysis leads us to conclude that GOHDE is effectively $\Lambda$CDM in disguise. This affirms our initial assumption that $\Lambda$CDM represents the best fit within the context of the GOHDE model. This conclusion is supported both analytically and through several statistical tests.

These considerations raise the question of whether HDE effectively addresses the issues associated with $\Lambda$CDM. It is important to note that this manuscript did not delve into the conceptual problems within HDE. However, it has become transparent that the construction of HDE has limitations and potential shortcomings. 

\section{Conclusions \label{sec:6}}

Let us summarize the findings presented in this article. We began with a quest to comprehend the characteristics of holographic dark energy using the Granda-Oliveros IR cut-off. In our pursuit of understanding these features, we discovered that Holographic Dark Energy (HDE) models fail to provide a precise solution despite their claim to address the cosmological constant problems. While they account for late-time acceleration, in GOHDE, the origin of the constant is shifted to an integration constant, which does not make an actual prediction. Thus, it leaves the old cosmological constant problem open.

The essence of any HDE model lies in the choice of the IR cut-off or the horizon entropy. Finding one with minimal issues is challenging despite various options in the literature. The Granda-Oliveros IR cut-off, which incorporates both the Hubble parameter and its derivative, closely resembles the Ricci scale in a flat universe and is considered an appropriate choice. Such a definition would fit perfectly with the total energy density but has less reason to suit dark energy alone.

This study demonstrates that HDE with the Granda-Oliveros IR cut-off is virtually the same as $\Lambda$CDM when we ignore radiation and curvature. While the free parameters can vary, the one that converges to $\Lambda$CDM is the most preferred. We explored the parameter space with various cosmological parameters to substantiate this argument. Although values different from the $\Lambda$CDM case can explain late-time acceleration, the model exhibits distinct behaviour during the past. The holographic energy tends to behave like the dominant energy component unless we carefully remove such traits. Like in the Ricci scalar, where $\alpha=2\beta$, the dark energy never behaves like radiation. However, this is not always the case. 

Another feature of the GOHDE is that the equation of state approaches zero during the matter-dominated era. This implies that the energy density, which behaves like dark energy, mimics matter in the early stages of cosmic evolution. This has no observational effects when we look only for scaling behaviours, just as we cannot distinguish between baryons and dark matter. However, it could significantly impact other early observations, where we can differentiate between them.

Further, the GOHDE density can be positive or negative, maintaining an overall conservation. Thus, the dark energy equation of state could become singular. The solution where dark energy does not transform into a matter-like entity corresponds to the $\Lambda$CDM case alone. While these behaviours do not violate the laws of physics, they may not accurately explain early-phase observations such as the CMB power spectrum, baryon density, etc. The $\Lambda$CDM is only recovered when we ignore radiation density, consistent with the current observations. The equation of state for dark energy provides further hints of a transition from early quintessence to late phantom nature and vice versa. This behaviour, uncommon among HDE models, leads to phantom crossing, which conflicts with the Generalized Second Law. Additionally, while exploring the features of the Hubble horizon, we observed that only the $\Lambda$CDM case adheres to the principle of entropy maximization. Further, age estimation also imposes stringent constraints on the values of free parameters, with the $\Lambda$CDM case being more desirable.

Regarding the choice of horizon entropy, we identify a crucial inconsistency in the formalism of HDE in general. Our analysis suggests that only the Hubble horizon gives a reasonable thermodynamic behaviour. It is unknown why the IR horizon does not, although it goes directly into the definition of HDE. These features challenge the very foundations of HDE, as we can trace the origin to the first law of horizon thermodynamics. 

We conducted a rigorous data analysis using $\chi^2$ minimization with the Markov Chain Monte Carlo method to substantiate our observations further. Our analysis indicates that the $\Lambda$CDM case provides the most natural best fits. When considering the full dataset, the best-fit value of $\beta$ with $w_{z_0}=-1$ is $0.6786^{+0.0095}_{-0.0097}$, which is statistically very close to $2/3$. When setting $w_{z_0}$ free, we find $\beta=0.686^{+0.048}_{-0.044}$ and $w_{z_0}=-0.988^{+0.042}_{-0.044}$ as the best fit under the Full data combination, which is again statistically close to $\Lambda$CDM. Thus, not only does the GOHDE model explain current observations, but it also supports the parameters that reduce it to the $\Lambda$CDM model.

To gain a better understanding of the model's significance, we explored various quantifiers such as $\chi^2$, $\chi^2_{\text{\tiny{dof}}}$, AIC, BIC, $\Delta$AIC, and $\Delta$BIC. Our estimates suggest that $\Lambda$CDM has an advantage, particularly when considering the $\Delta$BIC estimate, which firmly penalizes models for using extra parameters. Furthermore, we observe a strong correlation between the values of $\beta$ and $w_{z_0}$, as indicated by the similarity in the $\Delta$BIC values between GOHDE and $w$CDM models. Finally, we calculated the Bayes factor between different models and found that observational data support $\Lambda$CDM and GOHDE equally. Thus, virtually $\Lambda$CDM does hold the upper hand both phenomenologically and observationally. Similar conclusions were derived based on fine structure  measurement from ESPRESSO Consortium \cite{PhysRevD.105.123507}.

Another important aspect of our analysis was the usage of the standard Friedmann equations. Remember we plugged our expression for HDE density into the first Friedmann equation. It may not be the appropriate way to construct a holographic dark energy. As addressed in \cite{Manoharan2023} and \cite{golanbari2020renyi}, the choice of entropy will modify the form of the first law and thus the Friedmann equations derived from it. Thus, different entropy implies different equations of motion. If the definition of entropy is such that $S\sim L_{\text{IR}}^2$, it is only natural to assume that the choice of IR cut-off $L_{\text{IR}}$ can also lead to similar results. This will be an interesting future work. The fact that the GOHDE density itself looks like the second Friedmann equation also questions the validity of conventional HDE applied to dark energy alone \cite{HassanVT_2022}. 

In this manuscript, we did not consider any interaction between the dark sectors, which could significantly impact the analysis and yield unexpected results. However, our knowledge of the choice of interaction term is limited compared to our understanding of the nature of dark energy. In a related study \cite{PhysRevD.106.043527}, a generic interaction term was examined with the simplest IR cut-off, showing that while it can explain late-time acceleration, it results in a CMB power spectrum significantly different from that of $\Lambda$CDM. Thus, interaction models may not be a fruitful choice. Their calculations support our claims of an improper matter-dominated era for HDE. In this regard, it will be interesting to investigate whether there are tensions at high redshift (QSO data for $z\sim 7$ and above) similar to what was addressed in \cite{Risaliti2019} and scrutinized in \cite{PhysRevD.102.123532}. From an analysis perspective, more subtle reasons for similar bias and tensions were addressed in \cite{colgain2023mcmc}. 

Dark energy dynamics will undoubtedly influence large-scale structure growth, offering a potential test for dark energy models. A varying dark energy equation of state will significantly impact as we enter the matter-dominated epoch. While numerical methods are available to explore these features, they often assume a smooth or relatively smooth dark energy. Given that the equation of state significantly deviates from -1 to that of some diffused matter, this condition may not hold during the matter-dominated era. Therefore, we must adequately consider matter and dark energy perturbations as done in \cite{PhysRevD.106.043527}. Investigating these is beyond the scope of this article and is left as a topic for future research. 

\begin{acknowledgements}
Special thanks to Titus K Mathew, N. Shaji and Sarath Nelleri for the comments on the manuscript and the discussions on dark energy models. Sincere gratitude to Rhine Kumar A. K. and the Nuclear Physics Group at CUSAT for their kind provision of computational resources. Furthermore, I acknowledge the financial support from CSIR-NET-JRF/SRF, Government of India, Grant No: 09 / 239(0558) / 2019-EMR-I.
\end{acknowledgements}

\appendix
\section{Including radiation to GOHDE\label{AppendixA}}

The article concentrates on the period when the universe transitions from being dominated by matter to dark energy and omitted the radiation component from Eq. (\ref{FullHubble}). Here, we graph the progression of density parameters on a logarithmic scale to capture how they evolve with different values for $\alpha$ and $\beta$ in the presence of radiation. We assumed $\alpha=1$, which should be observationally identical to $\Lambda$CDM when we only consider the scaling behaviour of the Hubble parameter.
\begin{figure}[h]
	\includegraphics[width=0.48\textwidth]{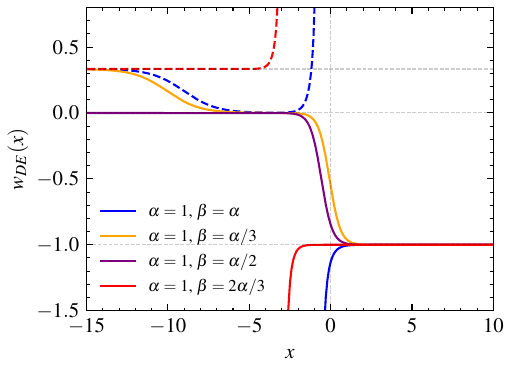}
	\caption{(colour online) Dark energy equation of state parameter plotted against $x=\log(a)$ for various relationships between $\alpha$ and $\beta$. The dashed line represents the equation of state when the dark energy density is negative.}
	\label{fig:withRadEoS}
\end{figure}

\begin{figure*}[t]
	\subfloat{\includegraphics[width=0.48\textwidth]{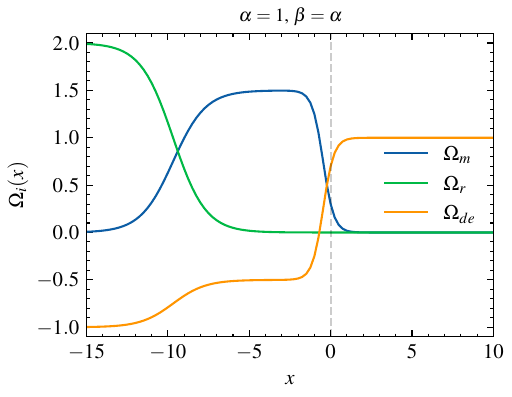}}
	\subfloat{\includegraphics[width=0.48\textwidth]{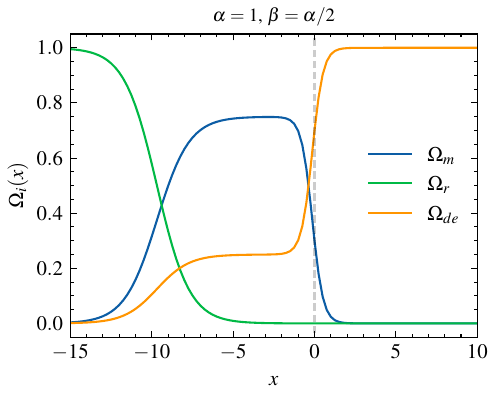}}\\
	\subfloat{\includegraphics[width=0.48\textwidth]{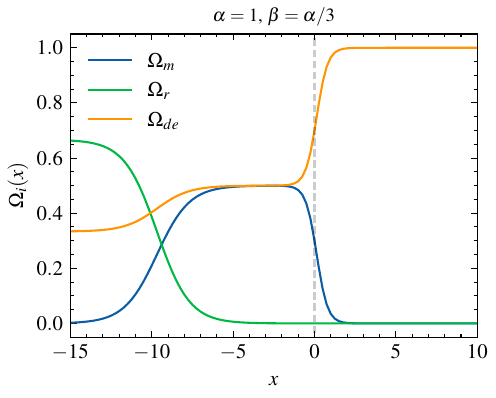}}\subfloat{\includegraphics[width=0.48\textwidth]{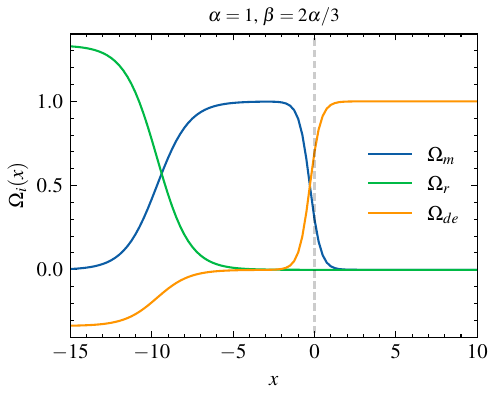}}
	\caption{(colour online) The evolution of matter density $\Omega_m$, radiation density $\Omega_r$, and dark energy density $\Omega_{de}$ are illustrated as a function of $x=\log(a)$ in the context of various relationships between $\alpha$ and $\beta$ within the GOHDE construction. The parameters are set to $\Omega_{\text{m0}}=0.3$ and $\Omega_{\text{r0}}=2.5\times10^{-5}$ for a flat universe.}
	\label{fig:densitywithrad}
\end{figure*}

For $\alpha=3\beta$, the dark energy equation of state parameter aligns with the effective equation of state. In this scenario, the scaling features of other fluids are retained for a flat universe. Consequently, the evolution of dark energy predominantly mirrors the behaviour of the dominant component, even when fixing the final de Sitter state by setting $\alpha=1$. However, the evolution of density parameters significantly diverges from the past considered in $\Lambda$CDM. In the case of $\alpha=3\beta/2$, as explicitly discussed in this article, the model closely resembles $\Lambda$CDM up to $a\sim 10^{-3}$, making it challenging for observational datasets to distinguish between them. Therefore, from an observational standpoint, excluding radiation has no significant impact. It will be interesting to extend these analysis to non flat scenario.

%

\end{document}